\documentclass[sigconf,nonacm]{acmart}

\usepackage{booktabs} 
\usepackage{multirow}
\usepackage{float}
\usepackage{subfig}

\newcommand{\one}{({\em i})\/}
\newcommand{\two}{({\em ii})\/}
\newcommand{\three}{({\em iii})\/}

\newcommand\st[1]{(\##1)\/}
\newcommand\s[1]{(\S\ref{s:#1})\/}

\usepackage{tabulary}
\usepackage[newcommands]{ragged2e}

\acmYear{2019}\copyrightyear{2019}
\setcopyright{rightsretained}
\acmConference[IoTDI '19]{IoTDI '19: Internet of Things Design and Implementation}{April 15--18, 2019}{Montreal, QC, Canada}
\acmBooktitle{IoTDI '19: Internet of Things Design and Implementation, April 15--18, 2019, Montreal, QC, Canada}
\acmPrice{15.00}
\acmDOI{10.1145/3302505.3310082}
\acmISBN{978-1-4503-6283-2/19/04}

\begin{document}
\title{
  Network Service Dependencies in
  \mbox{Commodity} Internet-of-Things Devices
}

\author{Poonam Yadav}
\affiliation{
 \institution{University of Cambridge}
 \streetaddress{Department of Computer Science \& Technology}
 \city{Cambridge, UK}
 \postcode{CB3 0FD}
}
\email{poonam.yadav@cl.cam.ac.uk}

\author{Qi Li}
\affiliation{
 \institution{University of Cambridge}
 \streetaddress{Department of Computer Science \& Technology}
 \city{Cambridge, UK}
 \postcode{CB3 0FD}
}
\email{ql272@cam.ac.uk}

\author{Anthony Brown}
\affiliation{
  \institution{University of Nottingham}
  \streetaddress{Horizon Digital Economy Research}
  \city{Nottingham, UK}
  \postcode{NG8 2BU}
}
\email{anthony.brown@nottingham.ac.uk}

\author{Richard Mortier}
\affiliation{
 \institution{University of Cambridge}
 \streetaddress{Department of Computer Science \& Technology}
 \city{Cambridge, UK}
 \postcode{CB3 0FD}
}
\email{richard.mortier@cl.cam.ac.uk}

\renewcommand{\shortauthors}{Yadav et al.}

%
%
%
%



\begin{abstract}
  We continue to see increasingly widespread deployment of IoT devices, with apparent intent to embed them in our built environment likely to accelerate if smart city and related programmes succeed. In this paper we are concerned with the ways in which current generation IoT devices are being designed in terms of their ill-considered dependencies on network connectivity and services. Our hope is to provide evidence that such dependencies need to be better thought through in design, and better documented in implementation so that those responsible for deploying these devices can be properly informed as to the impact of device deployment (at scale) on infrastructure resilience. We believe this will be particularly relevant as we feel that commodity IoT devices are likely to be commonly used to retrofit ``smart'' capabilities to existing buildings, particularly domestic buildings.

  To the existing body of work on network-level behaviour of IoT devices, we add \one~a protocol-level breakdown and analysis of periodicity, \two~an exploration of the service and infrastructure dependencies that will implicitly be taken in ``smart'' environments when IoT devices are deployed, and \three~examination of the robustness of device operation when connectivity is disrupted. We find that many devices make use of services distributed across the planet and thus appear dependent on the global network infrastructure even when carrying out purely local actions. Some devices cease to operate properly without network connectivity (even where their behaviour appears, on the face of it, to require only local information, e.g.,~the Hive thermostat). Further, they exhibit quite different network behaviours, typically involving significantly more traffic and possibly use of otherwise unobserved protocols, when connectivity is recovered after some disruption.
\end{abstract}

\maketitle

\section{Introduction}

It is widely believed that the number of Internet of Things (IoT) devices is growing rapidly and will exceed 20 billion by 2020~\cite{Gartner2017, BI2018, IHS2017}. A large part of this future growth is expected to come from sensors, actuators and computation deployed in the built environment. Governments, commercial organisations, and private citizens are all experimenting with how IoT devices can make us, our cities, and our infrastructure more efficient. Standards such as Building Information Modelling (BIM) level 2 are increasingly widely used to model building design and construction, and future iterations (anticipated BIM levels 3 and 4) are expected to cover development of the built environment and associated infrastructure (e.g.,~transport, refuse, utilities, communications, health, education) more broadly~\cite{BIM}. For example, the UK Government has required ``fully collaborative 3D BIM (with all project and asset information, documentation and data being electronic) as a minimum'' since 2016.\footnote{UK Government BIM level 2 mandate, \url{http://bim-level2.org/en/faqs/}} Digitisation of our built infrastructure looks set to accelerate.

We focus here on domestic contexts: smart homes where typical IoT devices might include environmental sensors, security cameras, personal health and wearable devices, voice controlled assistants, and robots. Other contexts seem likely to contain distinct but overlapping devices. For example, smart hospitals might have rather more wearable health monitoring and medical devices (drug monitoring and delivery systems, pacemakers, etc.), and might perhaps integrate smart medical robots to provide efficient end-to-end workflow in the hospital~\cite{Zhang2018}. In contrast, smart offices may share versions of a number of smart home devices for environmental sensing and control, while  adding systems targeting the shared workspace to provide features such as online resource and space booking. Smart cities will include all of the above and will add various infrastructure sensing and control systems for lighting, parking, refuse collection, traffic control, and so on. All involve both local and centralised processing of information with complex data and information flow among heterogeneous components, implying dependencies on a wide range of network services and protocols~\cite{Giang2016}.

The implications of this increased digitisation are not entirely clear however. For example, recent data breaches have continued to increase sensitivity to the potential impact on security and privacy of widespread sensing, and a number of authors have examined the network bandwidth implications of widespread IoT deployment. In this paper we focus on a slightly different question: what are the implications on our built environment in terms of resilience and robustness if we come to rely on Internet-connected IoT devices? We begin to address this question by analysing data collected from lab-based measurement of the behaviour and service dependency of a range of domestic IoT devices covering different application domains, manufacturers, and popularity~\s{method}. We analyse the collected data to understand traffic patterns, protocol usage, and service dependencies for the devices we monitor~\s{analysis}. We then look specifically at the robustness of these devices under network disruption, examining how they respond to different types of interruption to their connectivity~\s{robust}. Finally we put our study in context~\s{related} and conclude~\s{conclusions}.

\section{Methodology}
\label{s:method}

To begin to understand the data types, rates, and traffic patterns caused by different IoT devices, we deployed a set of commodity off-the-self IoT devices in a small test area in an office in our lab, and captured the Wi-Fi traffic generated by these devices. Other occupants of the office were notified that the devices were present, and we carefully did not analyse the data captured from the devices for anything other than its gross network characteristics. The data captured thus represents a ``minimum'' level of traffic, and so service dependency, as the devices were not interacted with as they might be in a more realistic deployment. Our aim in this work is to uncover baseline data about network and service dependency of a selection of devices rather than to study device behaviour ``in the wild''. We certainly hope to explore device behaviour of more devices in more realistic deployment scenarios in the future but that is not the subject of this paper.

\begin{table*}
  \centering
  \begin{tabulary}{\textwidth}{
      L L C L L C R R
    }
    \midrule
       & Device
       & {\bf H}ub/ {\bf S}ensor
       & Link Type
       & Protocols
       & {\bf S}ecure/ {\bf I}nsecure
       & Energy (W)
       & Avg. \mbox{Bandwidth} (B/s)
    \\

    \midrule
    1  & Hive Starter Kit Hub~\cite{Hive2018}              & H
       & Ethernet
       & TCP, IGMP, ICMP                                   & S
       & 1.8                                               & 120\\

    2  & TP-link Smart Plug~\cite{Tp2018}                  & H
       & Wi-Fi
       & UDP, TCP                                          & S,I
       & 2.05                                              & 100\\

    3  & Google Home Mini~\cite{Ghome2018}                 & H
       & Wi-Fi
       & UDP, TCP, IGMP, ICMP                              & S,I
       & 1.4                                               & 125\\

    4  & Amazon Echo Dot~\cite{Echo2018}                   & H
       & Wi-Fi
       & UDP, TCP, ICMP                                    & S,I
       & 1.95                                              & 125 \\

    5  & Arlo Security Camera Base Station~\cite{Arlo2018} & H
       & Ethernet, Wi-Fi
       & UDP, TCP                                          & S,I
       & 4.6                                               & 70\\

    6  & Foobot Air Quality Monitor~\cite{Foobot2018}      & S
       & Wi-Fi
       & TCP                                               & S
       & 1.79                                              & 18\\

    7  & Nest Smoke Alarm~\cite{Nest2018}                  & S
       & Wi-Fi
       & UDP, TCP                                          & S,I
       & NA                                                & 0.02\\

    8  & D-Link Motion Sensor~\cite{Dlink2018}             & S
       & Wi-Fi
       & UDP, TCP, IGMP                                    & S,I
       & 1.4                                               & NA\\

    9  & Hive Motion Sensor~\cite{Hivemotion2018}          & S
       & Zigbee                                            & HA 1.2
       & S                                                 & Battery & NA\\

    10 & ParrotPot Smart Flower Pot~\cite{Parrotpot2018}  & S
       & Bluetooth                                         & V4.0 BLE
       & S                                                 & Battery & NA\\

    11 & MiBand Smart Bracelet~\cite{Miband2018}           & S
       & Bluetooth                                         & V4.0
       & S                                                 & Battery & NA\\

    12 & Smart Bluetooth Tracker~\cite{Stracker2018}       & S
       & Bluetooth                                         & V4.0
       & S                                                 & Battery & NA\\
    \midrule
  \end{tabulary}
  \caption{\label{tab:DeviceList}IoT devices and their traffic behaviour summary.
  }
\end{table*}

\begin{figure}
  \includegraphics[width=\columnwidth]
  {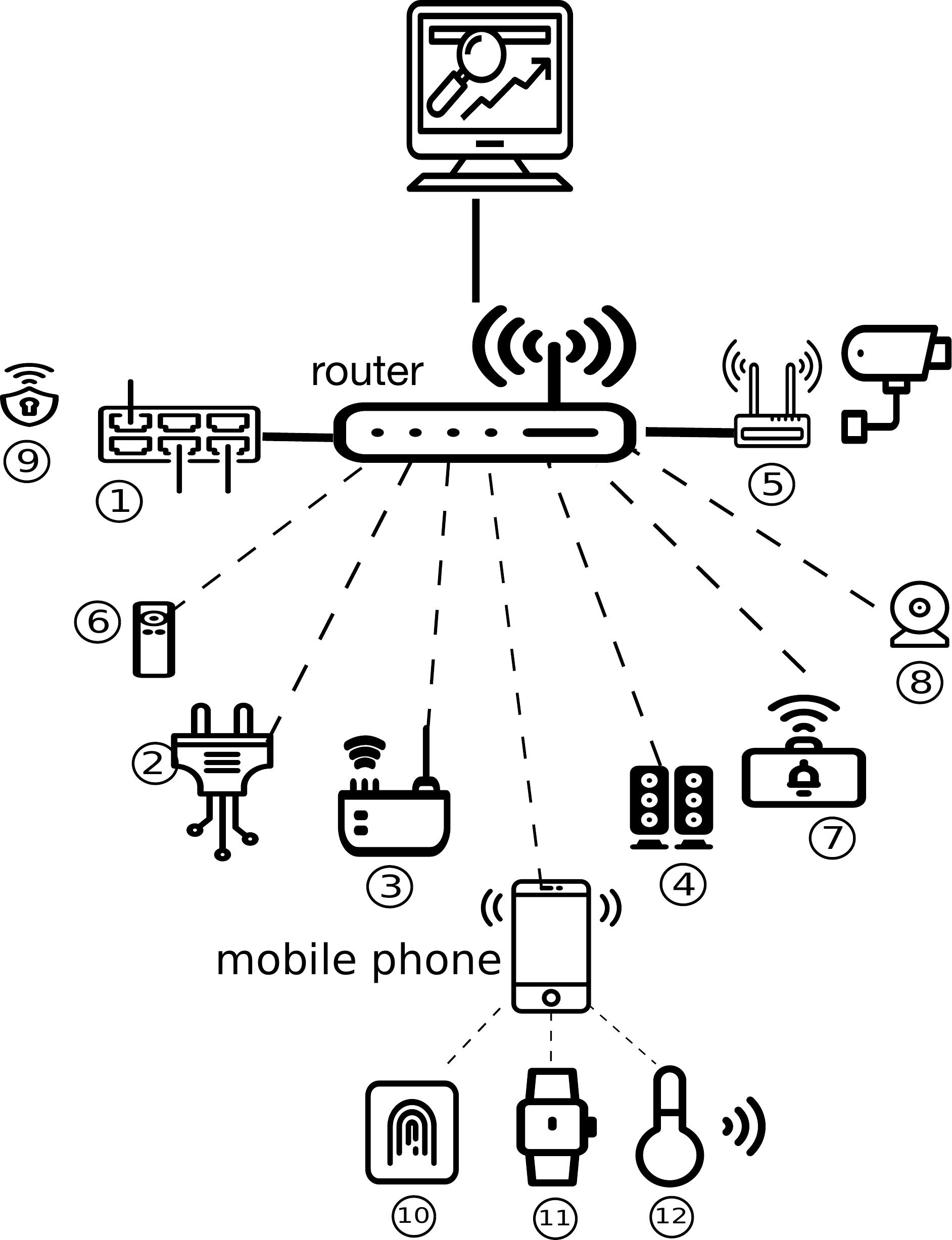}
  \caption{\label{setup}The Hive Hub (1) and Arlo Security Camera Hub (5) connect via wired Ethernet; the Hive Motion Sensor (9) communicates with the Hive Hub using Zigbee, and the Security Camera connects over Wi-Fi to the Arlo Security Camera Hub. Devices 10, 11, 12 connect to the smart phone over Bluetooth, and the rest (2, 3, 4, 6, 7, 8 and the controlling smart phone) connect over Wi-Fi. The router reaches the Internet via a wired Ethernet connection to the University's network. Detailed device descriptions are given in Table~\ref{tab:DeviceList}. }
\end{figure}

Table~\ref{tab:DeviceList} describes the commodity IoT devices we deployed. All were connected to a local Netgear N600 Wireless Dual Band Router WNDR3700v2 running Linux OpenWrt version 2.6.39.4~\cite{Netgear2018} either wirelessly over standard 802.11b Wi-Fi or via an Ethernet cable, allowing us to capture all traffic to and from each device. Figure~\ref{setup} depicts the experimental setup. We also made a very simple measurement of their energy consumption by connecting each device to a TP-Link Smart Plug for a fixed interval and recording the mean power consumption.

We categorise each device in one of two categories: \one~\textbf{Hub} refers to IoT devices which discover and control other IoT devices; \two~\textbf{Sensor} refers to IoT devices which connect to the router and then communicate directly with various cloud services without using any Hub.

To collect data from all IoT devices in our network, we ran \textit{monitor} and \textit{collect} scripts on the router. We first get a list of MAC addresses of the devices. On the router, we run the \textit{monitor} script on the interface that provides Wi-Fi access to the devices, filtering the traffic based on the MAC addresses of interest using \textit{tcpdump}~\cite{tcpdump2018}. On the other side, we schedule a cron job to periodically upload collected data to a local development machine for offline processing. This was necessary both because processing on the router would have been too slow given its limited processing capacity, and because the router has limited persistent storage, less than 50\,MB. Finally, we analyse all the \textit{pcap} files, focusing on packet headers and control protocols such as DNS.
For detailed packet analysis, we used Wireshark~\cite{Wireshark2018}, GraphViz~\cite{Graphviz2018} and a set of custom Python scripts.

The result was a dataset spanning 8th March, 2018 to 11th April, 2018 although, due to an unobserved device failure, the D-Link Motion Sensor ceased interacting over HTTPS after just one day, and ceased network activity altogether after just less than one week; hence we only have data for that device from 4th April, 2018 to 10th April, 2018.

\section{Analysis}
\label{s:analysis}

There are many different ways to understand the network communications behaviour of devices. In this section, we present analyses of the traffic we collected from the setup shown in Figure~\ref{setup}, as that traffic was transmitted and received by the devices listed in Table~\ref{tab:DeviceList}. Our purpose is not to make generalised statements about all IoT devices but to illustrate some of the ways commodity devices behave and to consider the implications of those behaviours. We particularly look at the implications for service dependency and device robustness in subsequent sections.

\subsection{Protocol Breakdown}

\begin{figure*}
  \centering
  \subfloat[Total traffic captured, by application.]{
    \label{serv_summary_total}
    \includegraphics[trim={0 0 0 .75cm},clip, width=0.33\textwidth]
    {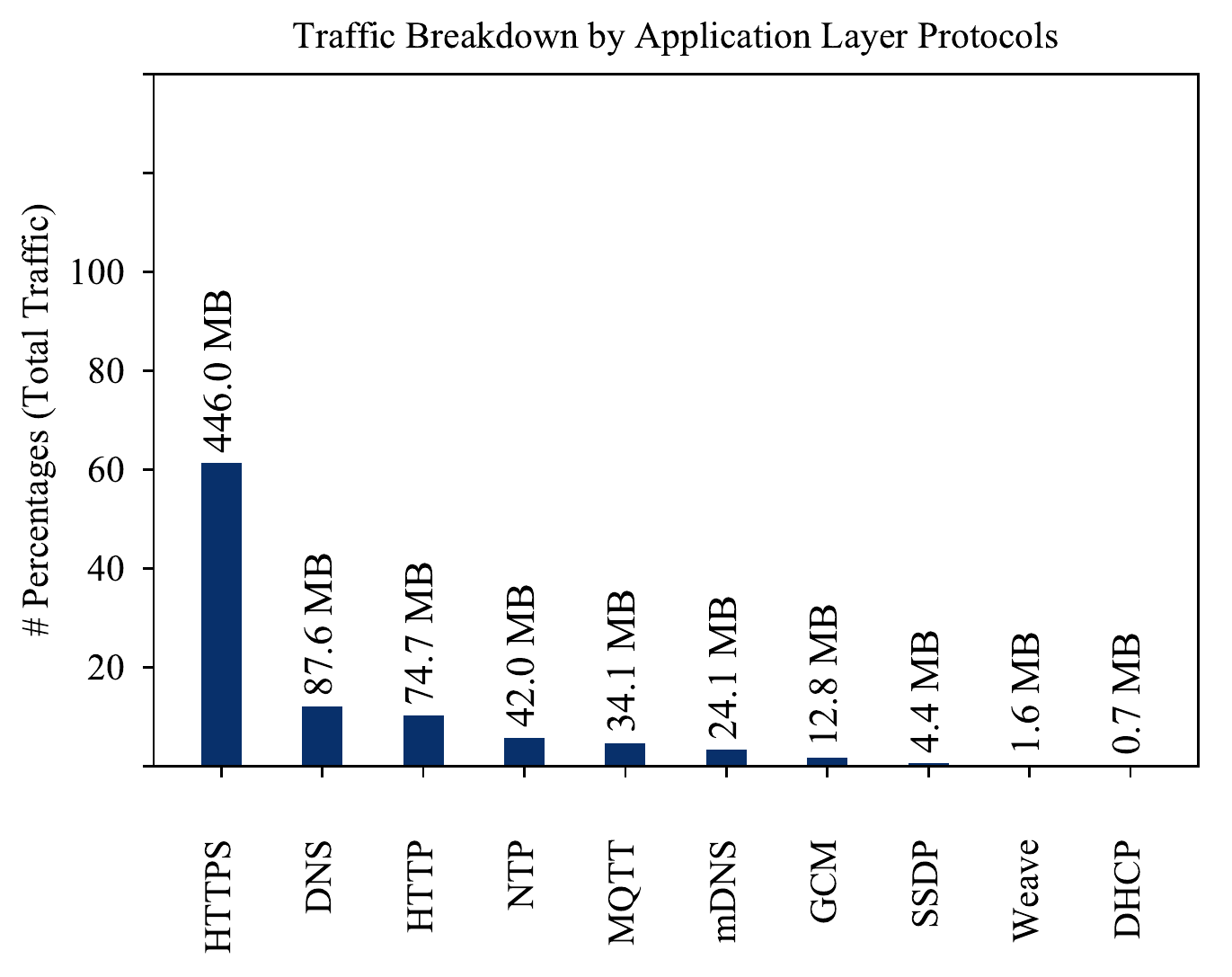}
  }
  ~
  \subfloat[Total traffic per device, by network protocol.]{
    \label{trans_summary}
    \includegraphics[trim={0 0 0 .75cm},clip, width=0.33\textwidth]
    {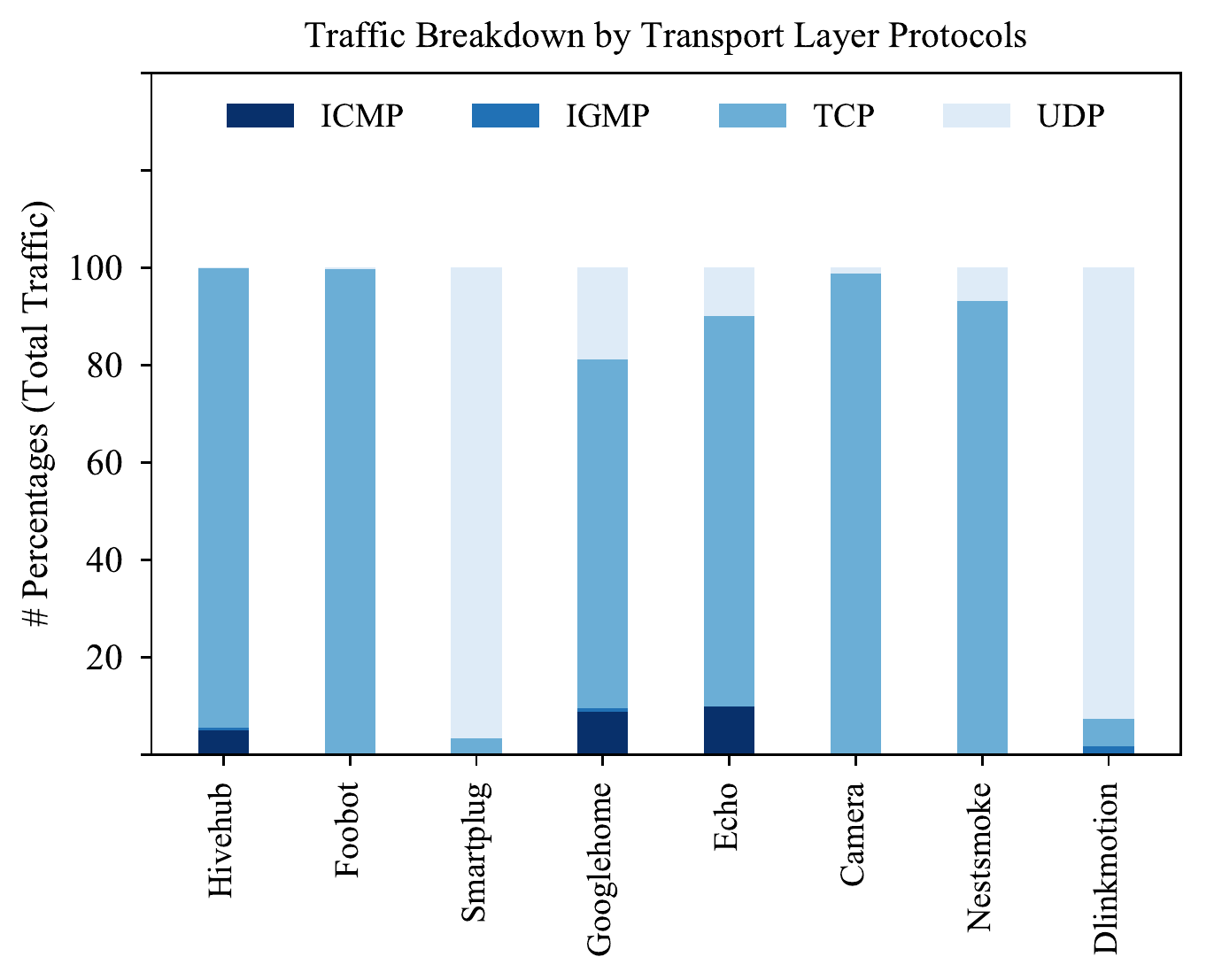}
  }
  ~
  \subfloat[Total traffic per device, by application.]{
    \label{serv_summary}
    \includegraphics[trim={0 0 0 .75cm},clip, width=0.33\textwidth]
    {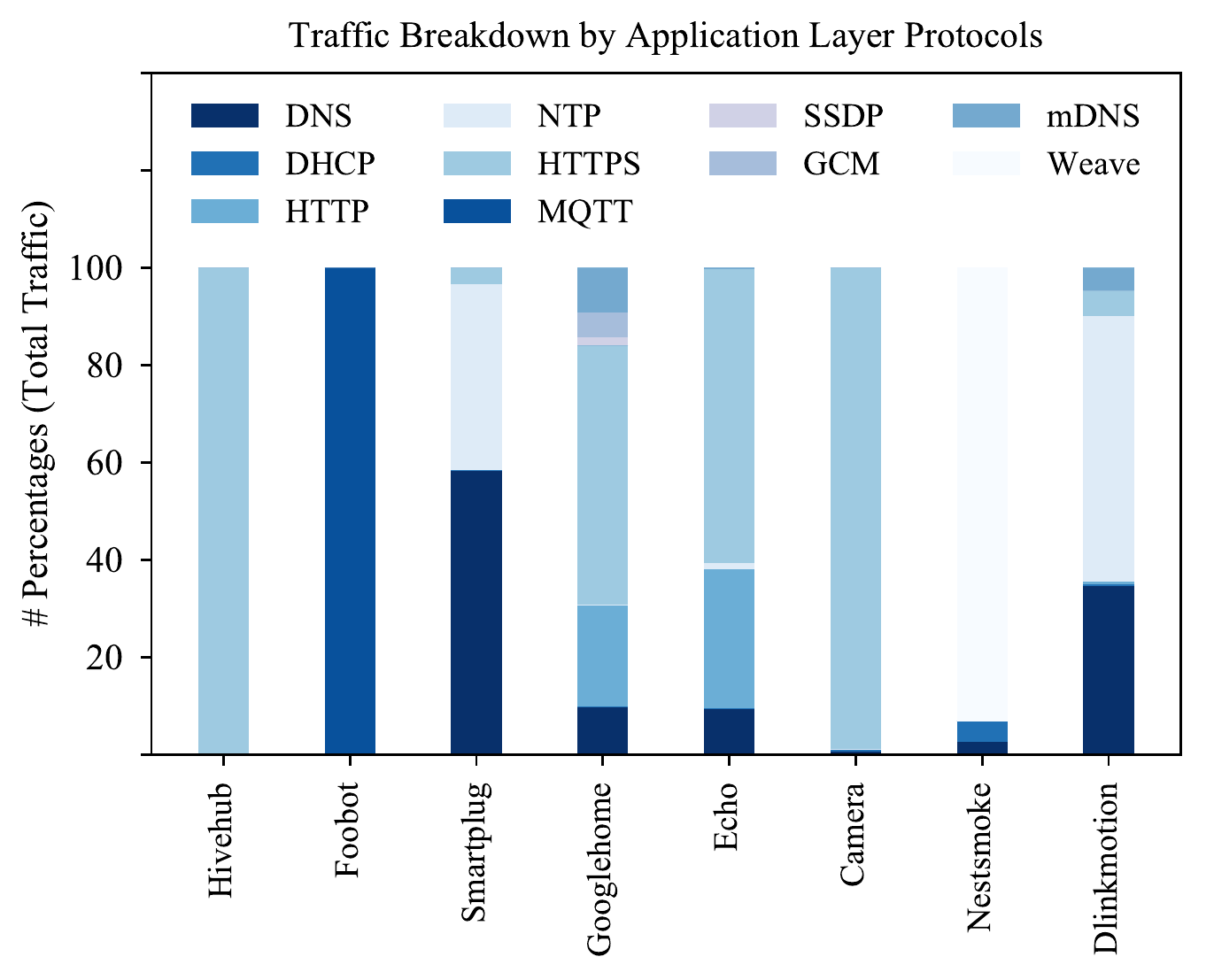}
  }
  \caption{\label{traffic_summaries}Traffic breakdown by protocol and device.}

\end{figure*}

Figure~\ref{traffic_summaries} presents a breakdown of the entire month's dataset by application protocol (Figure~\ref{serv_summary_total}), and per device by network and application protocol (Figures~\ref{trans_summary} and~\ref{serv_summary}). It is surprising to observe how much NTP, DNS and mDNS is in use by two devices in particular, the Smart Plug and D-Link Motion Sensor. It is also interesting to observe that only one device makes significant use of a classical IoT protocol (MQTT, used by the Foobot), though the Nest device also uses an IoT specific protocol (Weave) that was proprietary until released into Nest's developer platform in 2015. The rest use standard web protocols such as HTTP and HTTPS.

\paragraph{Local Network}

For pairing and device discovery, many IoT hubs use low power and short range communication protocols to connect to devices (sensors). These protocols include Zigbee (IEEE 802.15.4)~\cite{Zigbee2018}, Lora~\cite{Lora2018}, Zwave~\cite{Zwave2018}, Lightwave~\cite{Lightwave2018}, Bluetooth~\cite{Bluetooth2018}, RFID communication (LF (125--134 kHz), HF (13.56 MHz), UHF (433, and 860--960 MHz))~\cite{Rfid2018}. In our setup we have a few devices directly connecting to Hub using Zigbee, Bluetooth and Wi-Fi, e.g.,~the Hive motion sensor connects to the Hive Hub using the Zigbee protocol. We have three sensor devices which communicate with smartphone apps using Bluetooth and then those apps communicate with cloud services over the smartphone's Wi-Fi connection via the router.

\paragraph{Encrypted Traffic}

One straightforward observation we can make from the collected traffic traces is to observe the use of secure communication protocols between IoT devices and the outside world~\cite{Gebski2006}. Figure~\ref{serv_summary} categorises traffic generated from each device based on application layer protocols. All IoT devices send at least part of their traffic using HTTPS, with \textbf{Hub} devices sending more ($>$50\%) compared to \textbf{Sensor} devices. However we have not yet investigated how secure is the use of HTTPS by different devices in terms of selection of appropriate encryption suites and TLS configurations.

\subsection{Traffic Characterisation}



\begin{figure*}
  \centering
  \subfloat[Hive Hub]{
    \label{fig:hivehub_summary_w}
    \includegraphics[width=.33\textwidth]{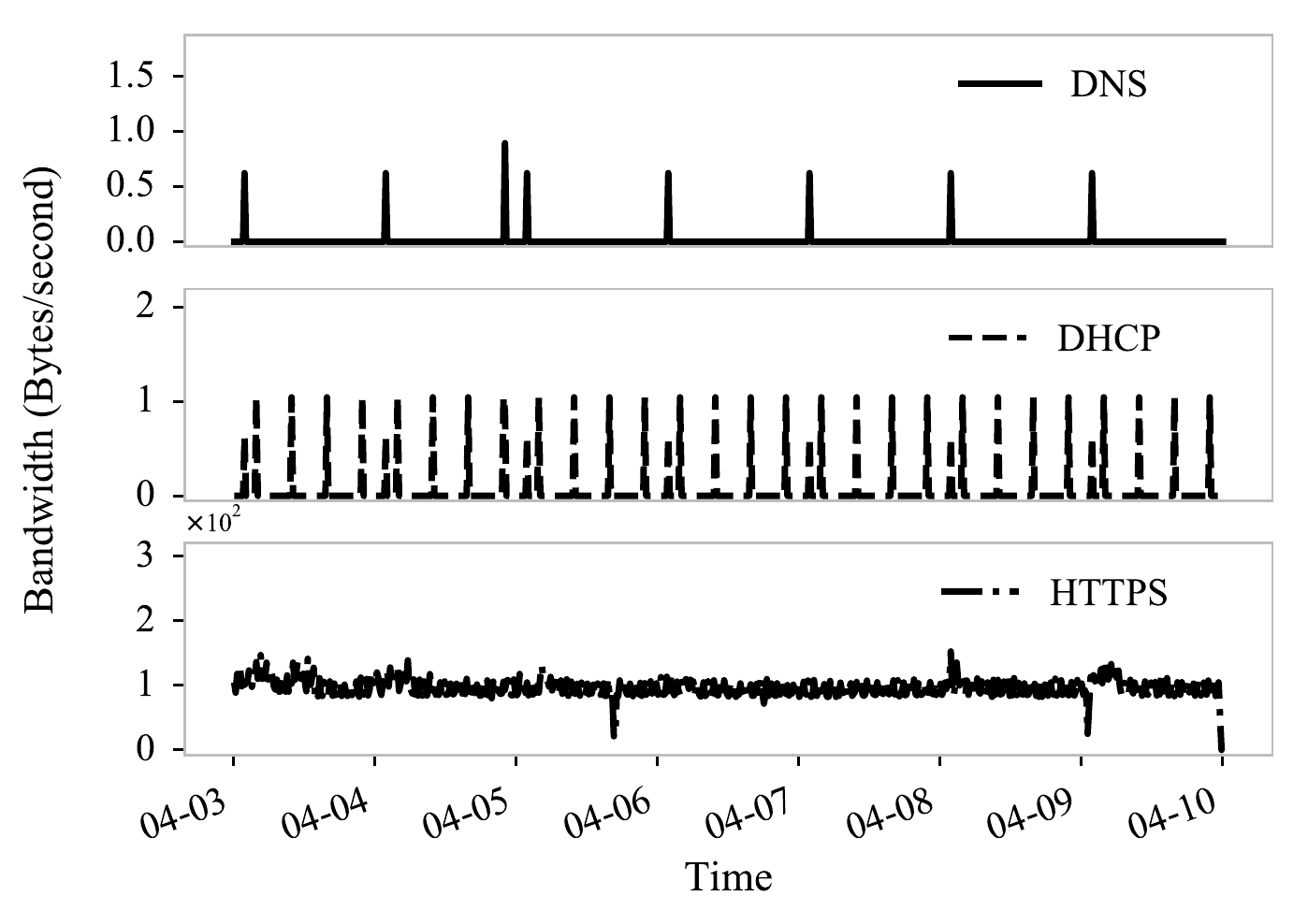}
  }
  ~
  \subfloat[Foobot]{
    \label{fig:foobot_summary_w}
    \includegraphics[width=.33\textwidth]{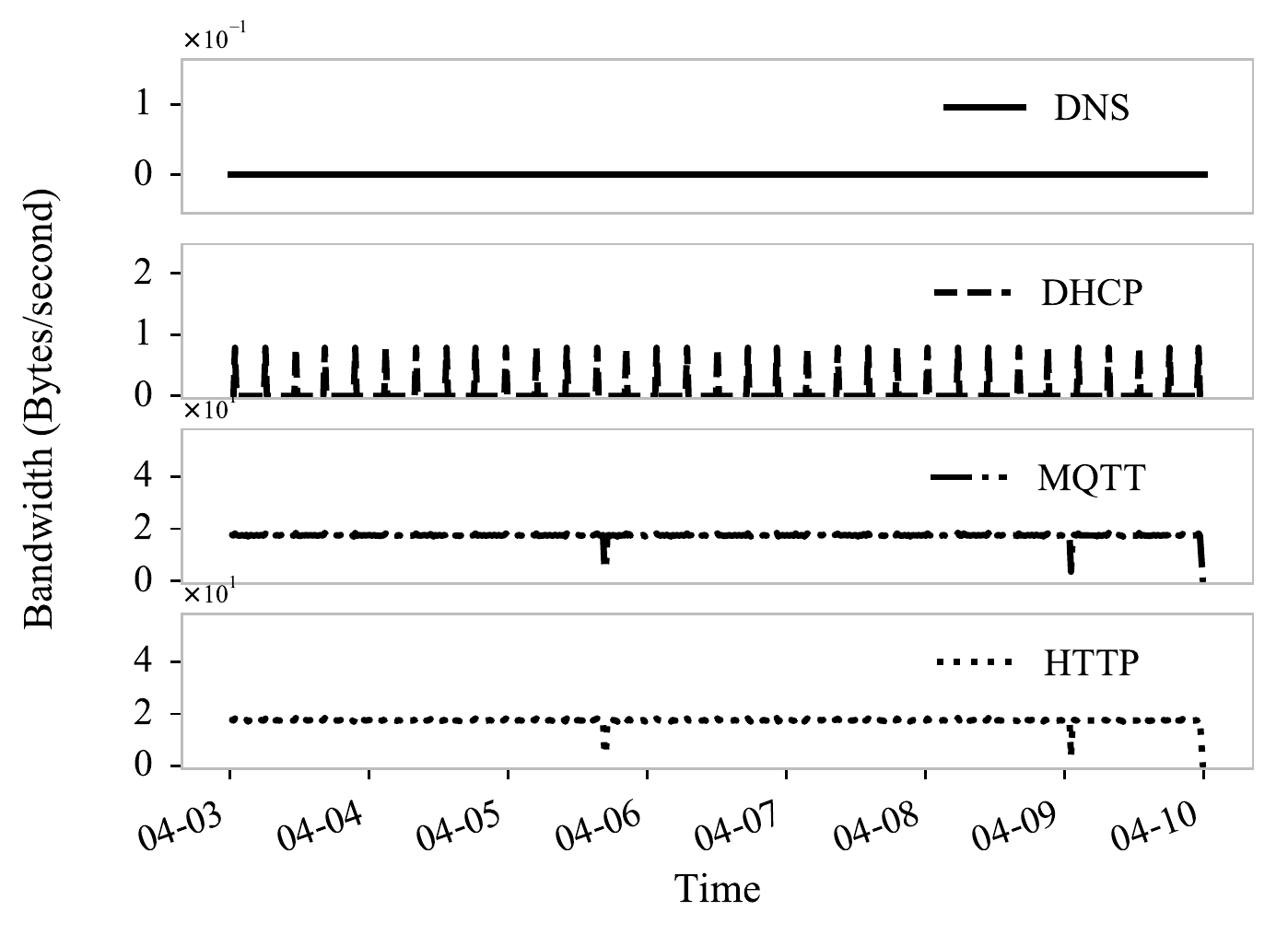}
  }
  ~
  \subfloat[Smart Plug]{
    \label{fig:smartplug_summary_w}
    \includegraphics[width=.33\textwidth]{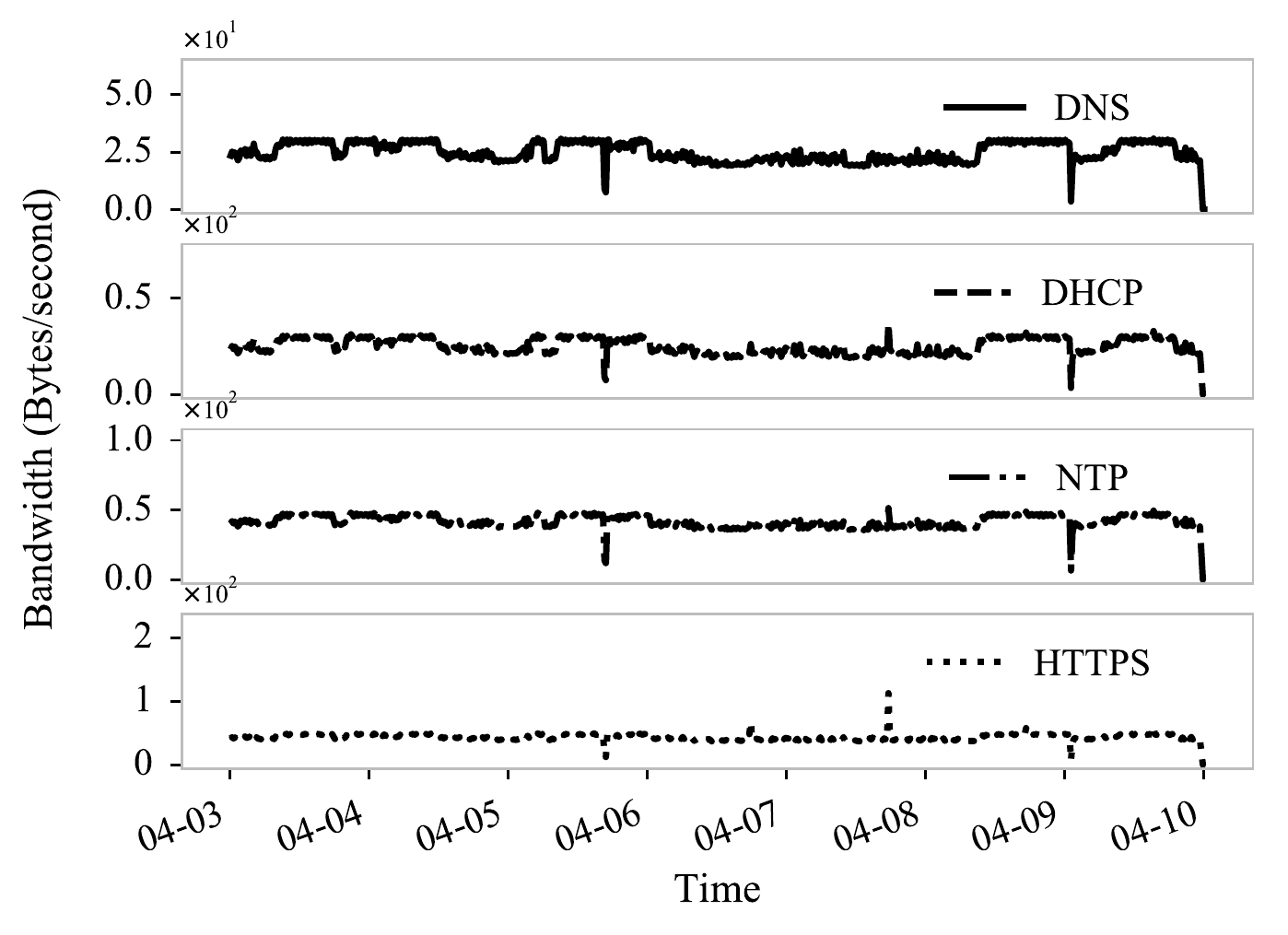}
  }
  \\

  \subfloat[Google Home]{
    \label{fig:googlehome_summary_w}
    \includegraphics[width=.33\textwidth]{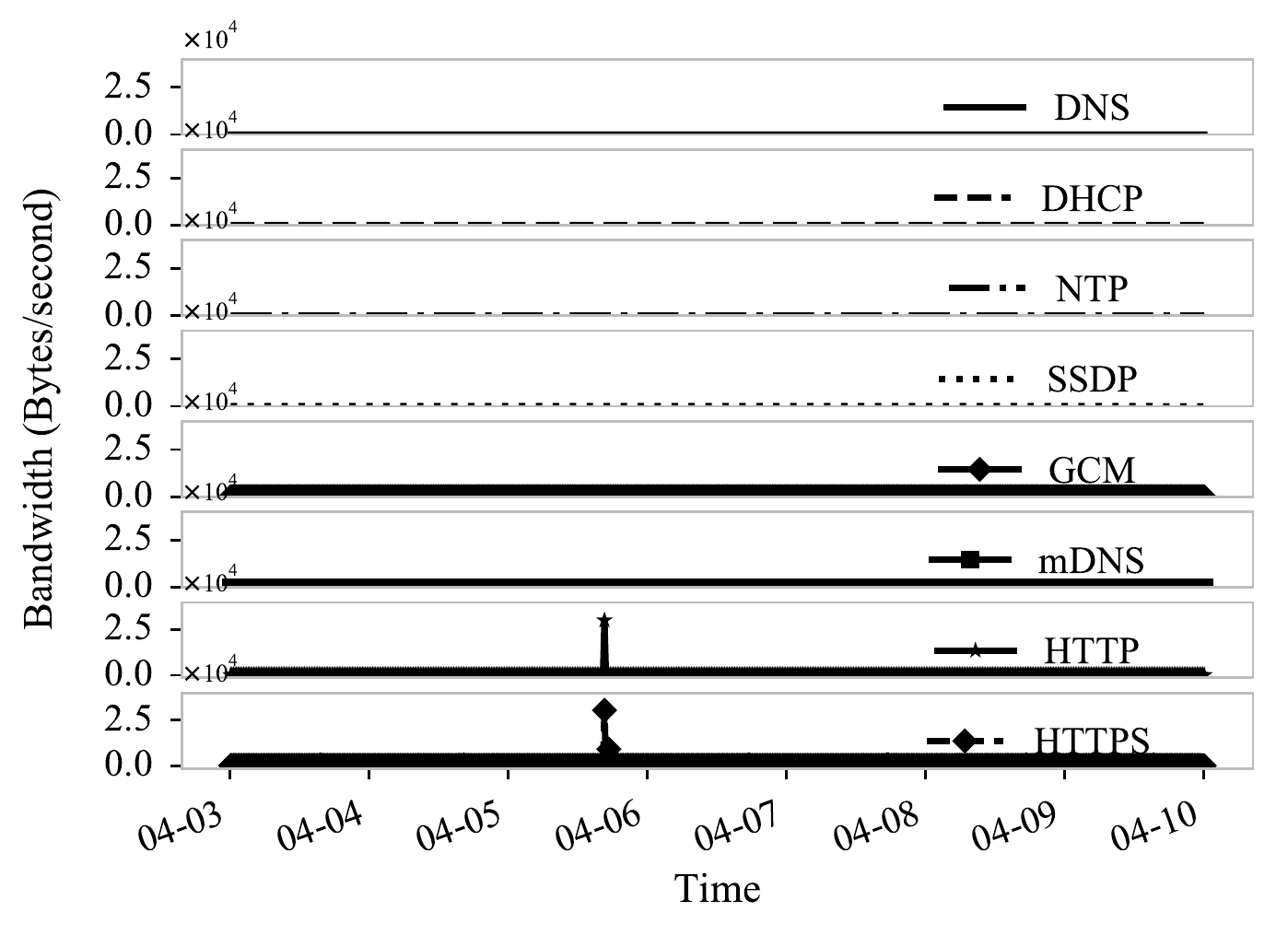}
  }
  ~
  \subfloat[Amazon Echo]{
    \label{fig:echo_summary_w}
    \includegraphics[width=.33\textwidth]{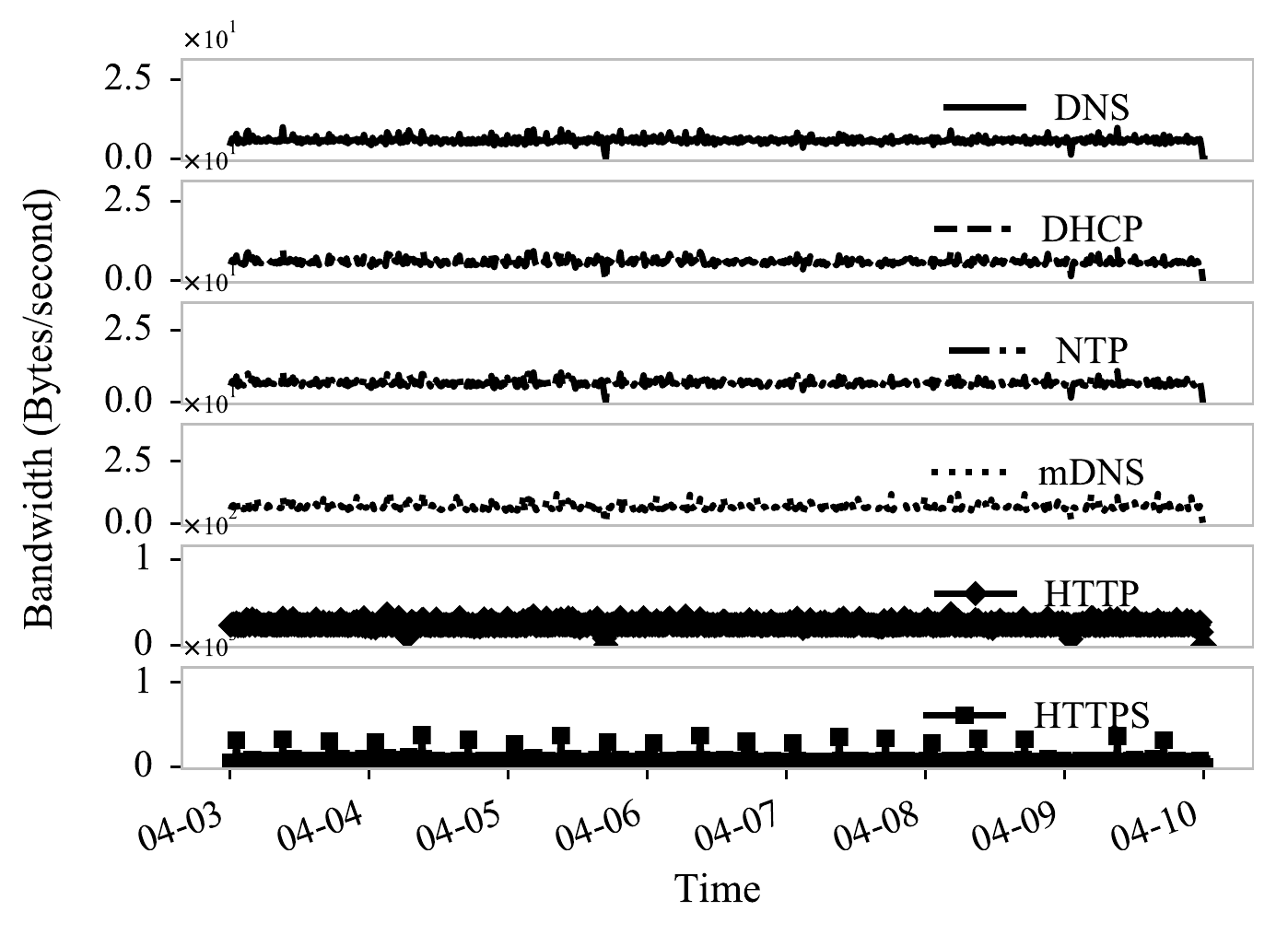}
  }
  ~
  \subfloat[Security Camera]{
    \label{fig:camera_summary_w}
    \includegraphics[width=.33\textwidth]{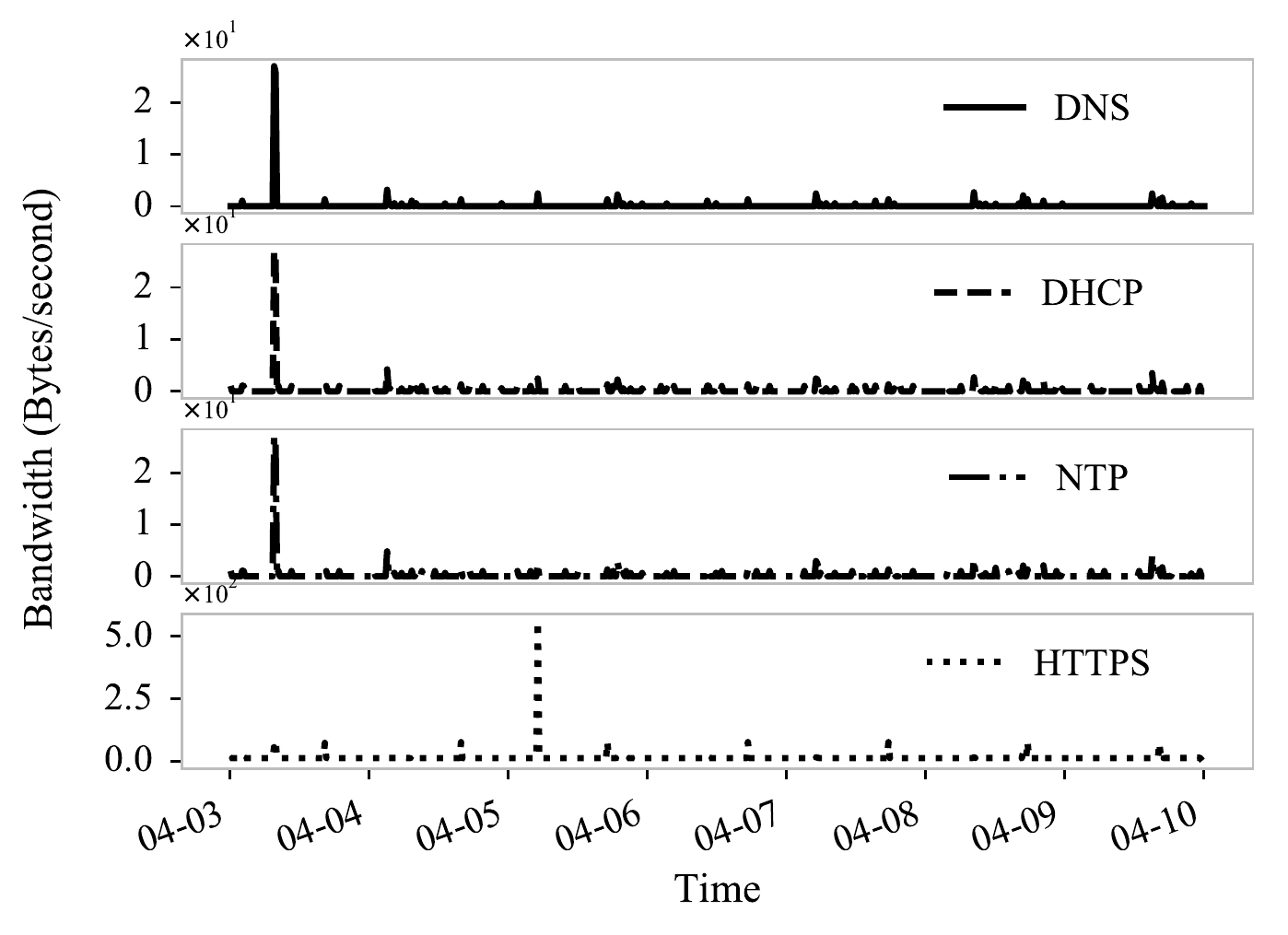}
  }
  \\

  \subfloat[Nest Smoke Alarm]{
    \label{fig:nestsmoke_summary_w}
    \includegraphics[width=.33\textwidth]{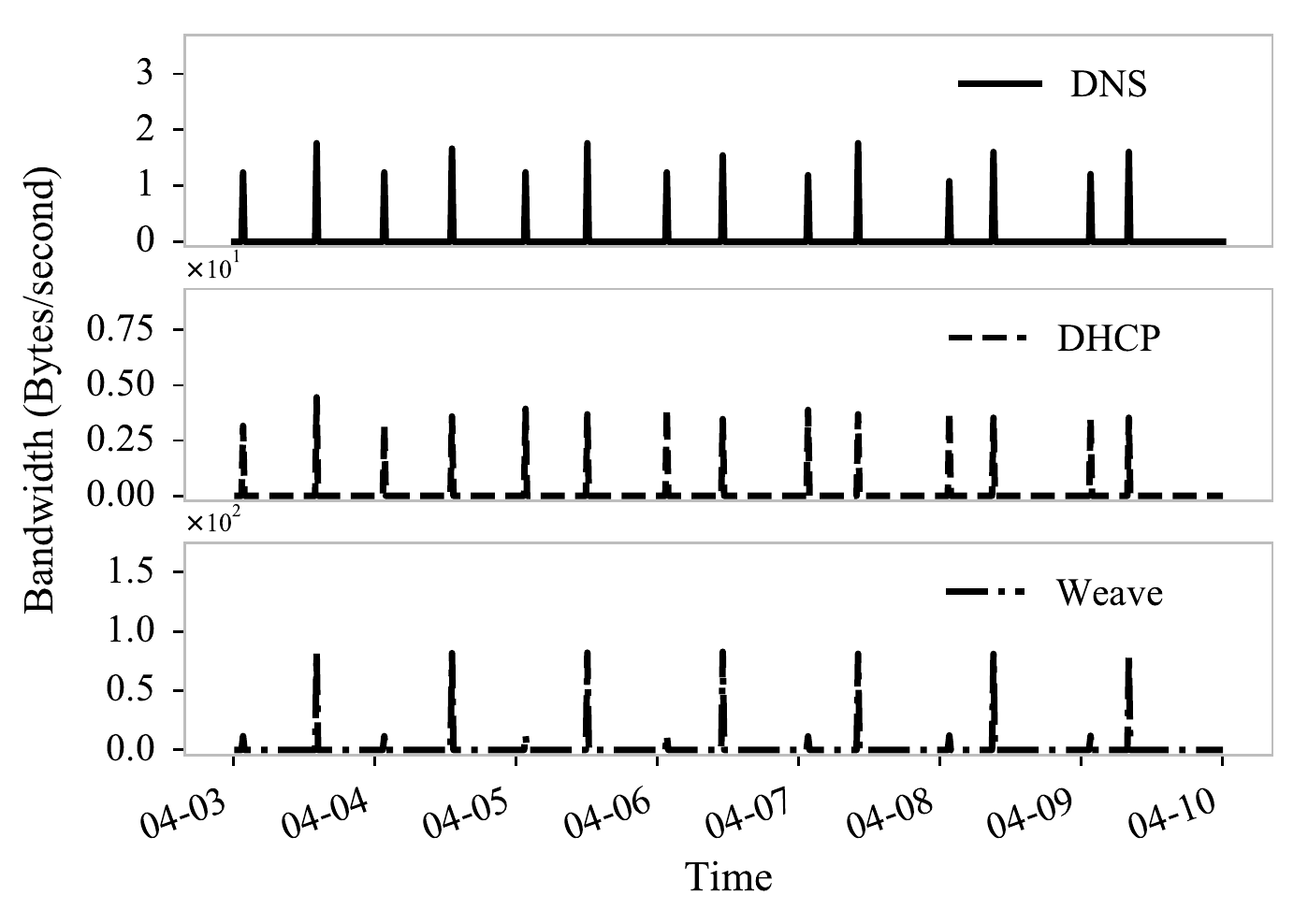}
  }
  ~
  \subfloat[D-Link Motion]{
    \label{fig:dlinkmotion_summary_w}
    \includegraphics[width=.33\textwidth]{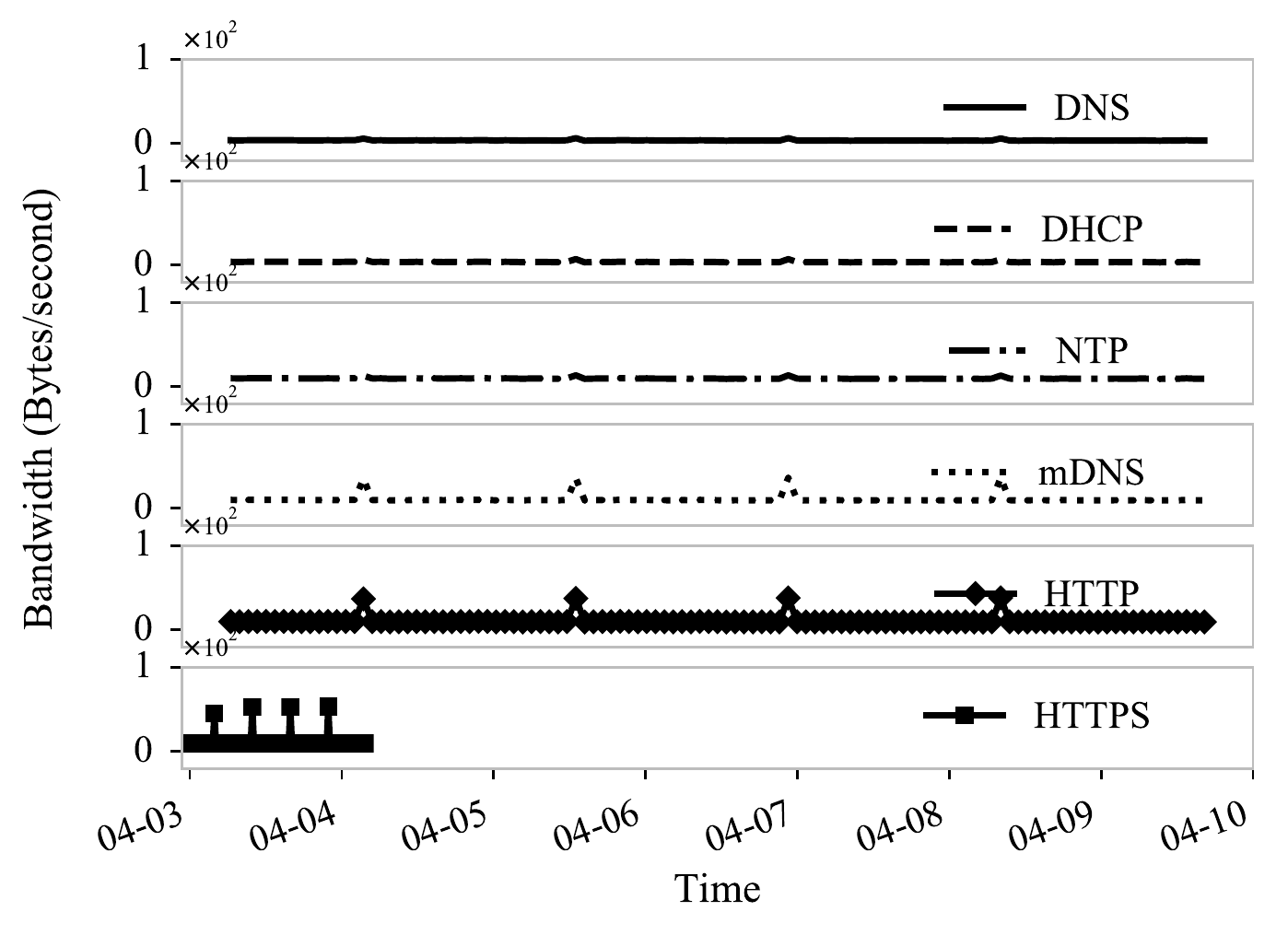}
  }
  \\
  \caption{\label{f:series_2}Per-device bandwidth used over one week, in 15\,minute buckets. Note variation in $y$-axis scales. As noted above, an unobserved device failure truncated the D-Link Motion trace after one day for HTTPS and completely after one week. }
\end{figure*}

Figure~\ref{f:series_2} shows bandwidth consumption by considering aggregate bytes transmitted and received in 15\,minute windows over one week.

The upward spikes shown for the Hive Hub (Figure~\ref{fig:hivehub_summary_w}), Smart Plug (Figure~\ref{fig:smartplug_summary_w}) and Security Camera (Figure~\ref{fig:camera_summary_w}) are caused by devices receiving software updates, while the downward spikes are due to network congestion at the local router due to a Google Home software update early in the morning of 6th April (Figure~\ref{fig:googlehome_summary_w}).

Amazon Echo (Figure~\ref{fig:echo_summary_w}), Nest Smoke Alarm (Figure~\ref{fig:nestsmoke_summary_w}) and D-Link Motion sensor (Figure~\ref{fig:dlinkmotion_summary_w}) showed some periodic upward spikes associated with frequent software updates. A high spike in the Security Camera Hub (Figure~\ref{fig:camera_summary_w}) is due to the camera image-capture triggered by its motion sensor. In summary, we found $>99\%$ of the Hive Hub's (Figure~\ref{fig:hivehub_summary_w}) and Security Camera Hub's (Figure~\ref{fig:camera_summary_w}) total traffic is composed of HTTPS packets and the remaining traffic comprises a few periodic DHCP, NTP and DNS interactions.

The majority of Foobot's (Figure~\ref{fig:foobot_summary_w}) traffic consists of MQTT running over TCP. Some of the devices, e.g.,~Smart Plug (Figure~\ref{fig:smartplug_summary_w}), Amazon Echo (Figure~\ref{fig:echo_summary_w}) and D-Link Motion sensor (Figure~\ref{fig:dlinkmotion_summary_w}), send frequent NTP traffic, and we see it forms a significant percentage of total traffic sent by these devices. As compared to all other devices, the traffic rate generated by Nest Smoke Alarm is small when it is in its ideal listening mode; it sends just 6 packets per day (totalling around 180 bytes per day). Nest Smoke Alarm uses the \textit{Weave} protocol over TCP to communicate twice a day to the Nest Cloud Service.

\begin{figure*}
  \centering
  \subfloat[Hive Hub]{
    \label{fig:hivehub_fft}
    \includegraphics[width=.33\textwidth]{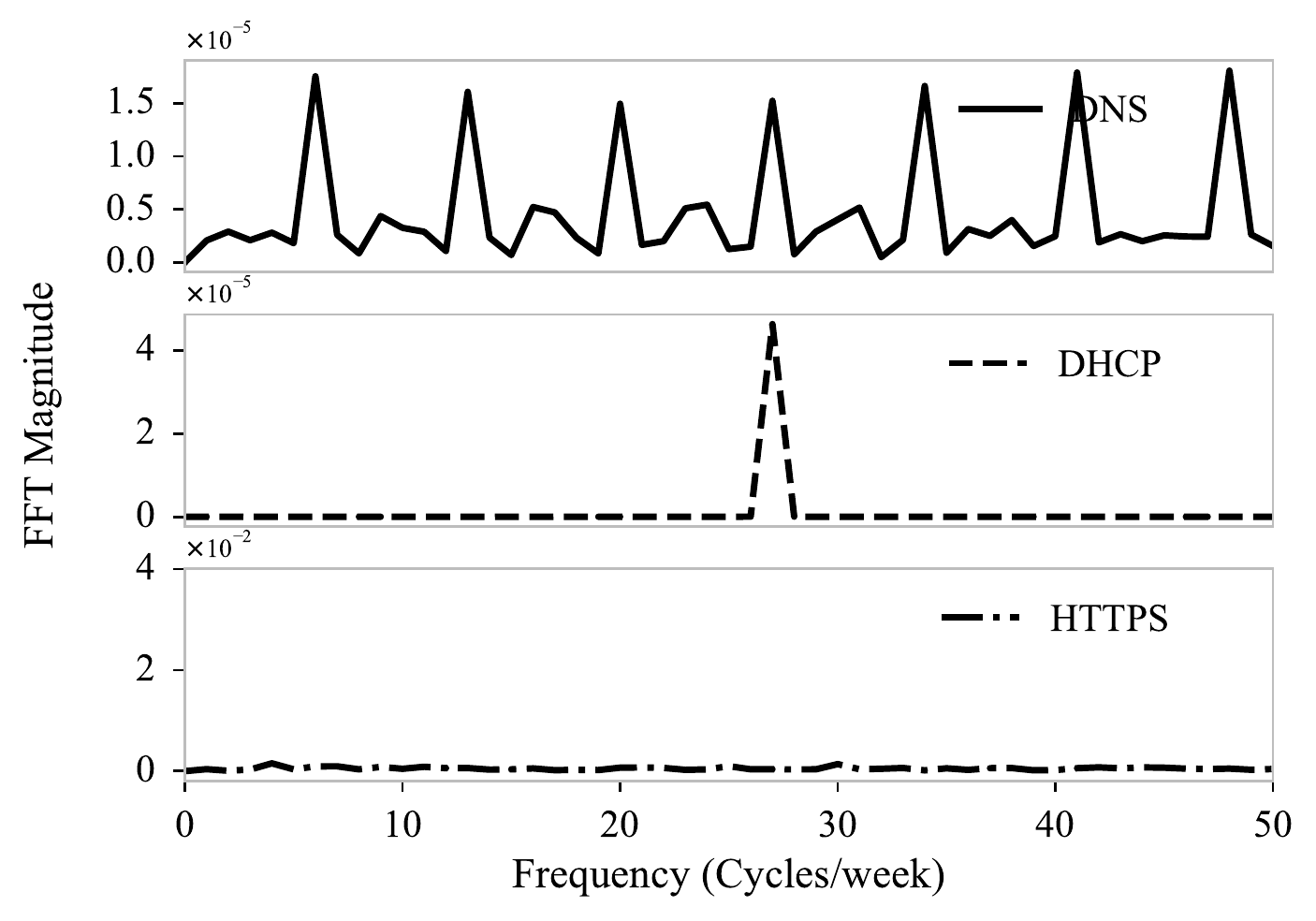}
  }
  ~
  \subfloat[Foobot]{
    \label{fig:foobot_fft}
    \includegraphics[width=.33\textwidth]{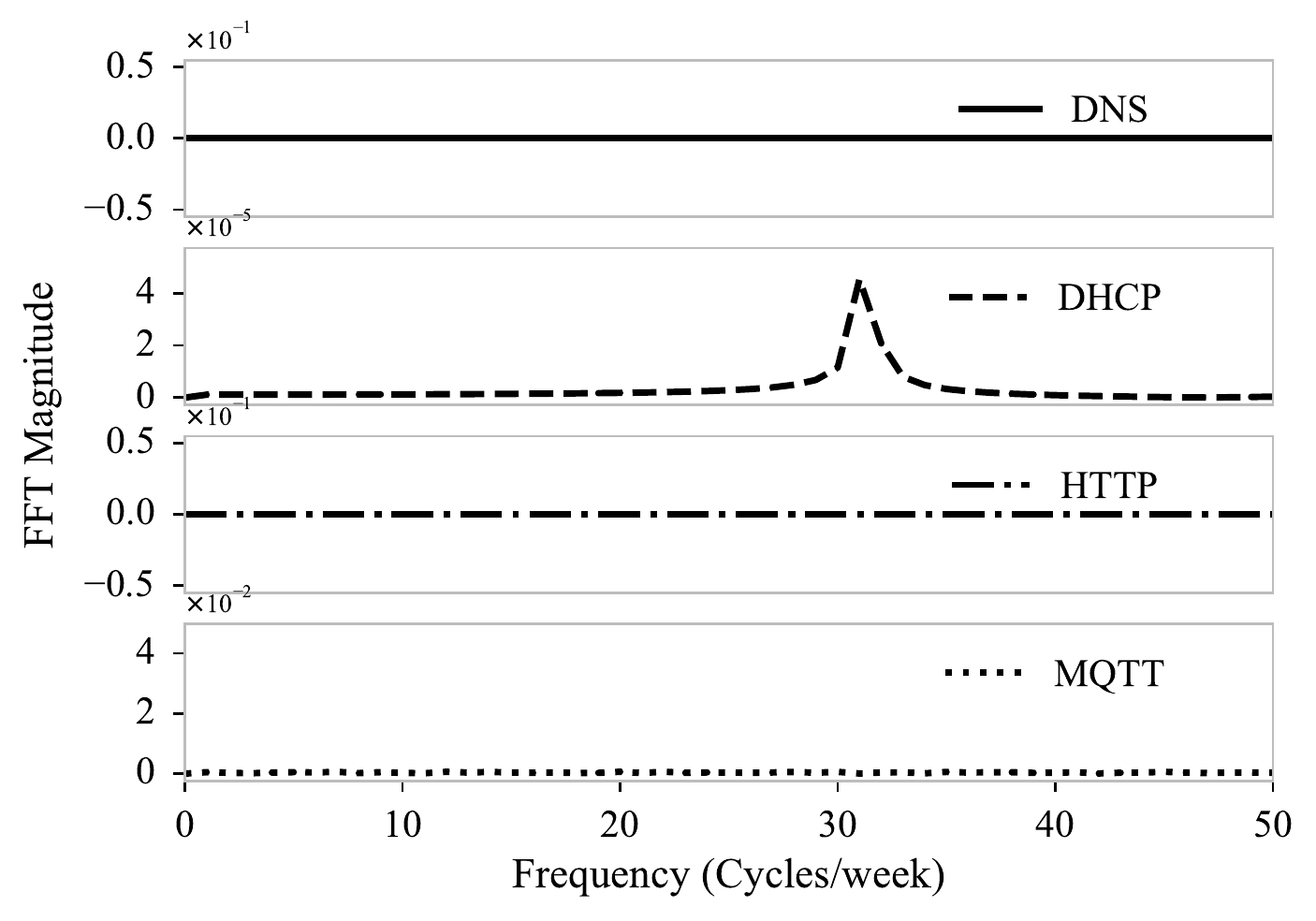}
  }
  ~
  \subfloat[Smart Plug]{
    \label{fig:smartplug_fft}
    \includegraphics[width=.33\textwidth]{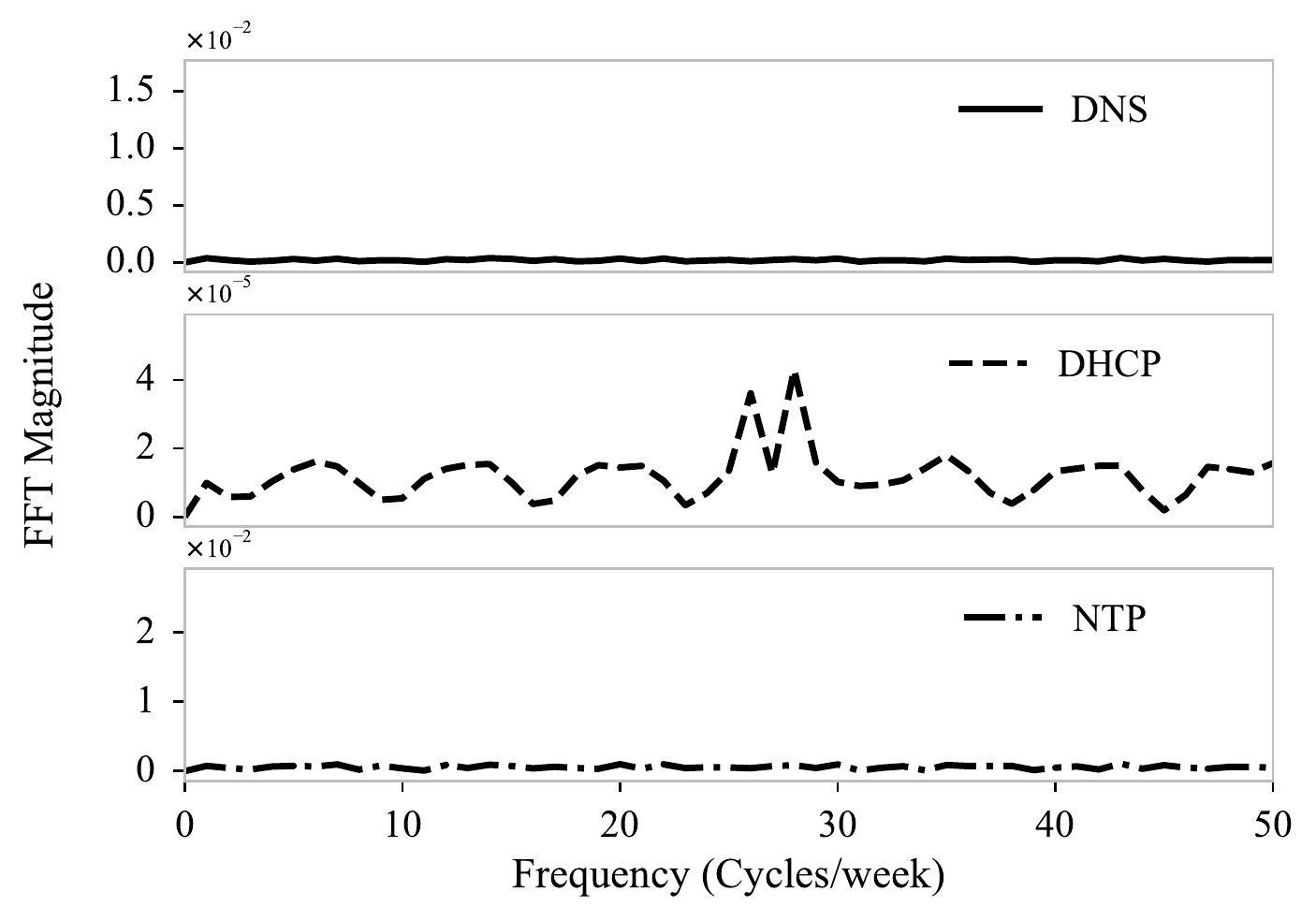}
  }
  \\

  \subfloat[Google Home]{
    \label{fig:googlehome_fft}
    \includegraphics[width=.33\textwidth]{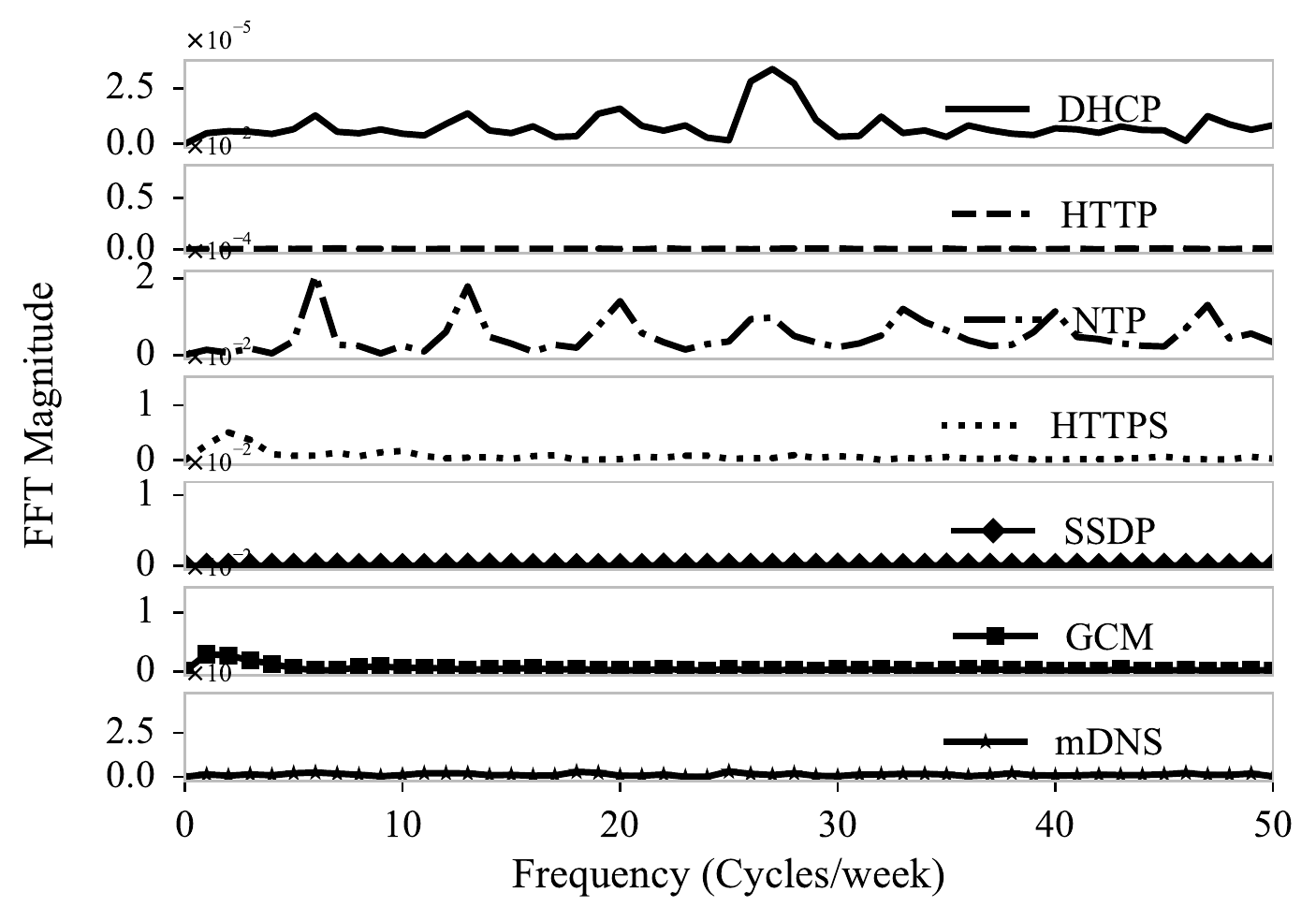}
  }
  ~
  \subfloat[Amazon Echo]{
    \label{fig:echo_fft}
    \includegraphics[width=.33\textwidth]{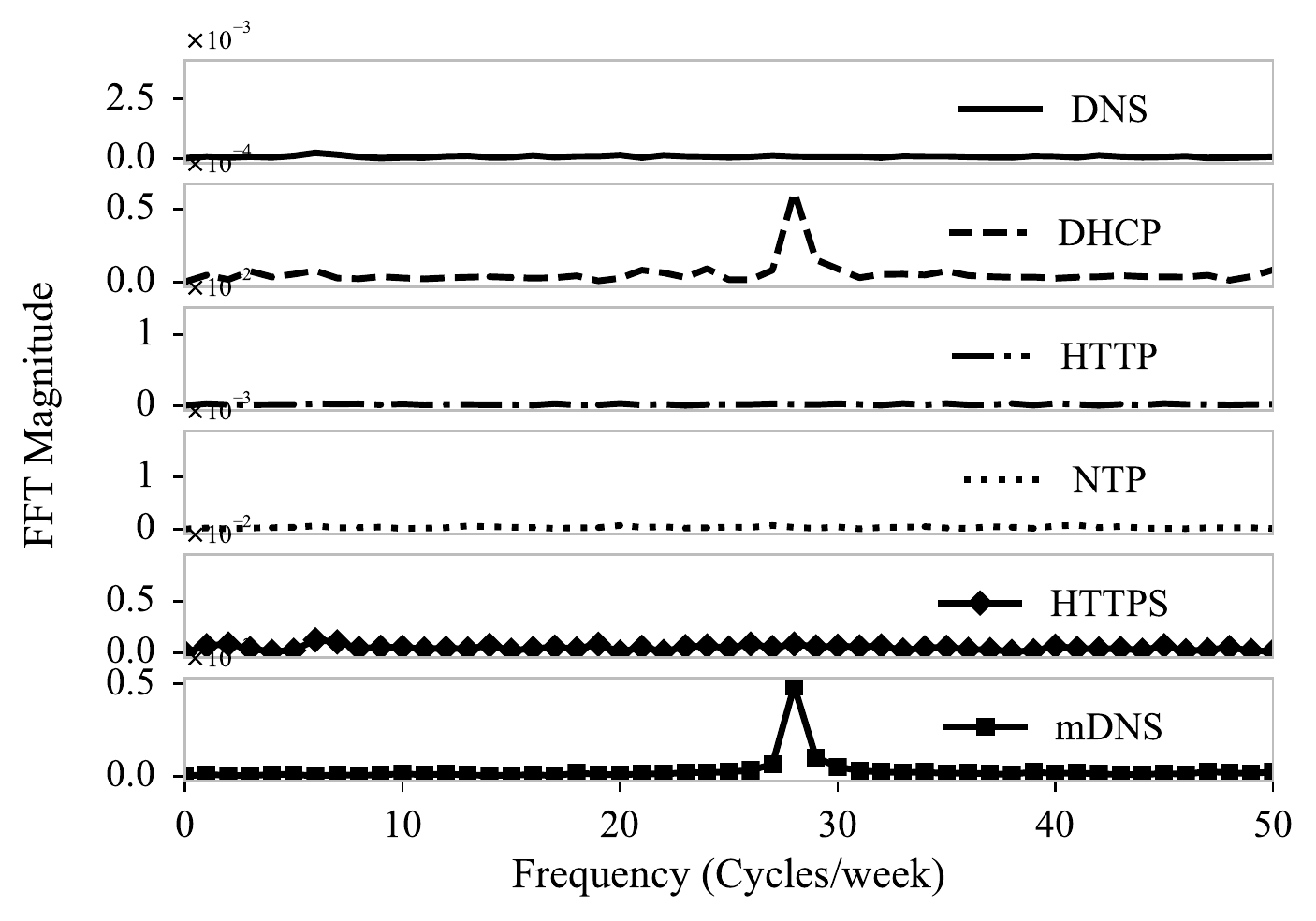}
  }
  ~
  \subfloat[Security Camera Hub]{
    \label{fig:camera_fft}
    \includegraphics[width=.33\textwidth]{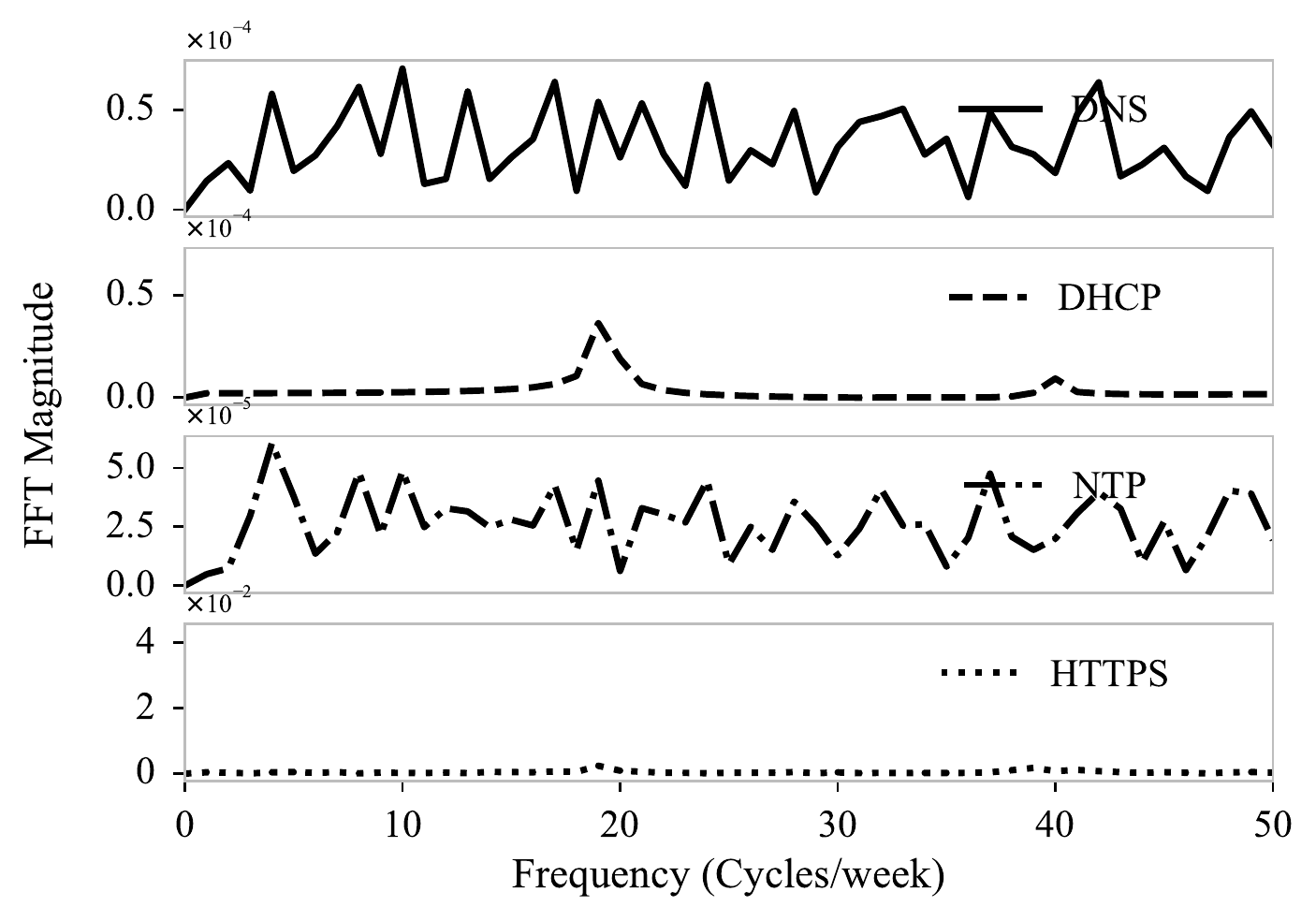}
  }
  \\

  \subfloat[Nest Smoke Alarm]{
    \label{fig:nestsmoke_fft}
    \includegraphics[width=.33\textwidth]{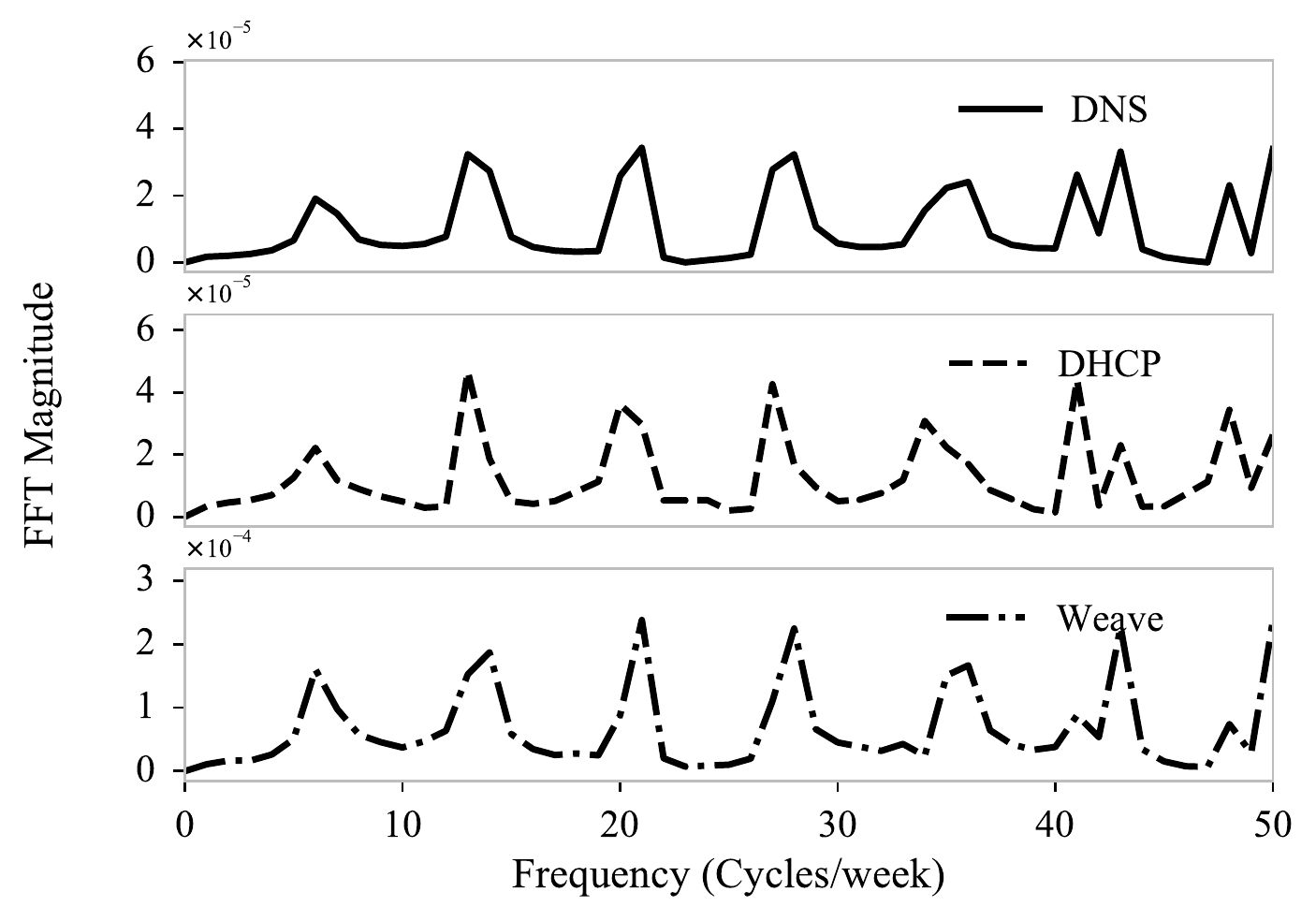}
  }
  ~
  \subfloat[D-Link Motion]{
    \label{fig:dlinkmotion_fft}
    \includegraphics[width=.33\textwidth]{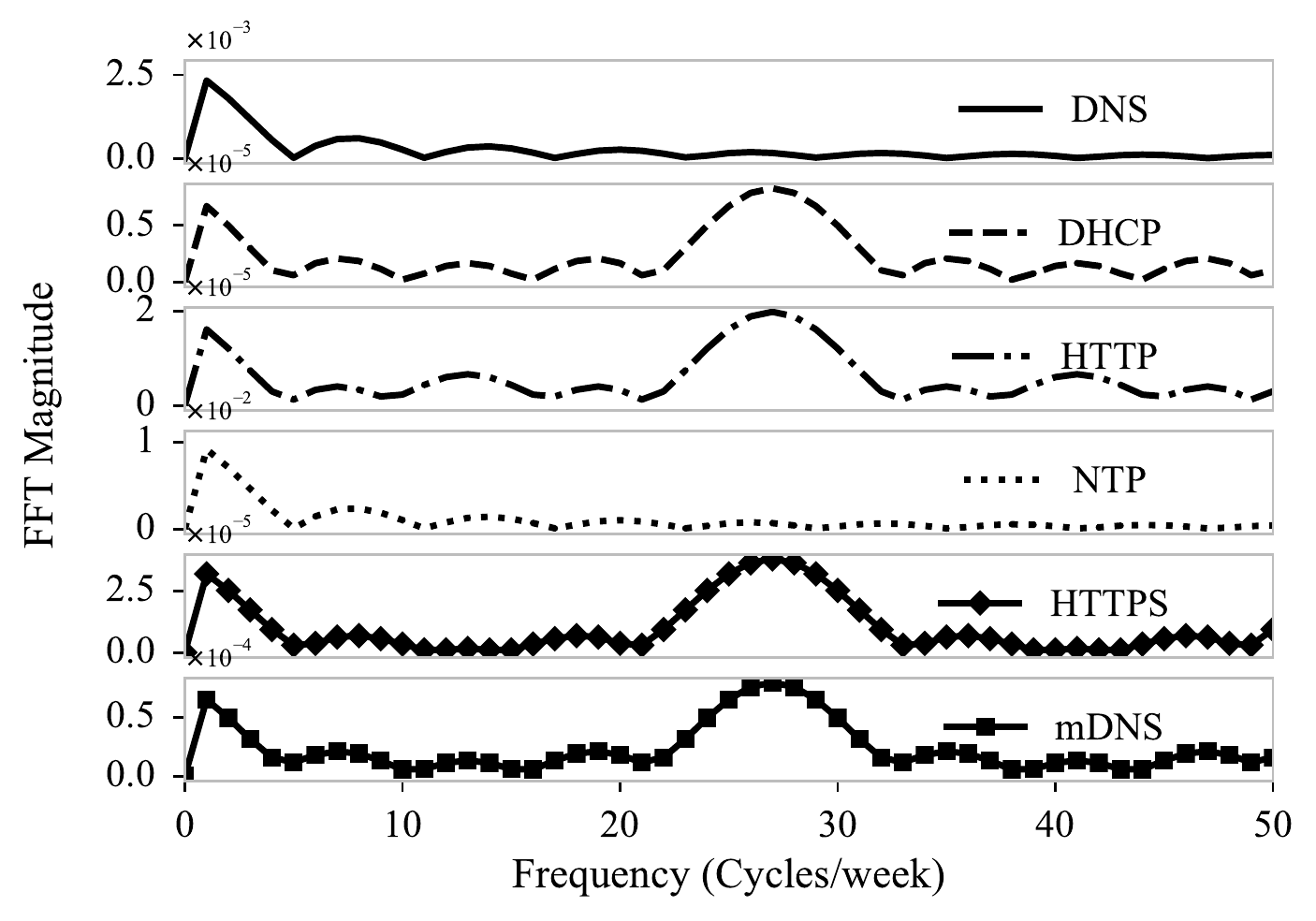}
  }
  \\
  \caption{\label{f:series}The periodicity of activity visible on the traffic generated for each device over one week by applying an FFT to the time-series data. }
\end{figure*}

We then analysed traffic from each device for periodicity across all active protocols during one week, 3rd April 2018 to 10th April 2018. We applied Discrete Frequency Fourier Transform~(FFT) to detect periodicity in network duty cycle, and we present normalised the FFT magnitude vs frequency (cycles/week) in Figure~\ref{f:series}. Our devices appear to all use DNS and DHCP periodically with other protocols typically used more intermittently.

We see that the Hive Hub uses only three application protocols (Figure~\ref{fig:hivehub_fft}) with periodicity of 28 cycles/week for DHCP traffic, an average of 4 DHCP requests every day. DNS traffic shows different periodic cycles 7, 14, 21, 28, 35, 42, 49 cycles/week. It means some DNS requests frequency were in the range of 1 request per day to 7 requests per day. The Foobot network usage is aperiodic for three protocols with only DHCP showing periodicity at 32 cycles/week, i.e.,~4--5 DHCP requests/day (Figure~\ref{fig:foobot_fft}). Out of three protocols used by Smart Plug, we see periodic behaviour with DHCP with two clear peaks, 24 and 28 cycles/week. However, there is some cyclic behaviour in DNS which means the device is making many DNS requests with periodicity 7, 14, 21, 35, 42 cycles/week as well (Figure~\ref{fig:smartplug_fft}). Google Home uses seven application layer protocols and found four of them show some periodic behaviour. There are continuous NTP requests with a periodicity of 7, 14, 21, 35, 42, 49 cycles/week. DHCP period is strongest at 28 cycles/week, but shows smaller peaks for 7, 14, 21, 35, 42 (Figure~\ref{fig:googlehome_fft}).

We also observed activity for GCM (Google Cloud Messaging) and HTTPS (analysed in the next section) at 1--2 cycles/week. Amazon Echo traffic composed of six protocols and DHCP and mDNS showed clear periodicity pattern at 28 cycles/week whereas HTTPS traffic showed periodicity 2--4 and 28 cycles/week in Figure~\ref{fig:echo_fft}. Security Camera Hub showed clear periodicity for DHCP 18 cycles /week and various frequency peaks in DNS and NTP traffic as shown in Figure~\ref{fig:camera_fft}. Nest Smoke Alarm showed periodic behaviour with all three protocols exhibiting frequencies 7, 14, 21, 28, 35, 42, 49 cycles/week (Figure~\ref{fig:nestsmoke_fft}). D-Link Motion sensor showed periodic behaviour at 1 and 28 cycles/week, (Figure~\ref{fig:dlinkmotion_fft}). However, as noted above, this device suffered some unnoticed failure after just one day resulting in HTTPS traffic ceasing to be observed.

\subsection{Protocol \& Service Dependency}

It is inevitable that IoT devices in deployment will depend on connectivity via a range of Internet protocols, both locally and to potentially many cloud-hosted services. We now examine some of these protocol and service dependencies for the devices we measured to illustrate some of the complex dependencies our infrastructure will take on if we increasingly deploy commodity IoT devices.

NTP usage in particular varied considerably between different devices, in terms of both the servers accessed and frequency. Some devices, e.g.,~the Hive Hub, used NTP during the setup process only. Some, e.g.,~the Foobot air quality monitor and Nest Smoke Alarm, use embedded timing protocols via MQTT and Weave rather than NTP. All those using NTP communicated with NTP servers run by the manufacturer except for the TP-Link SmartPlug which made extensive use of the global NTP Pool project servers~\cite{Tpntp2018}.


\begin{figure}
  \centering
  \subfloat[Hive Hub]{
    \label{fig:hivehub_geo}
    \includegraphics[height=22mm]{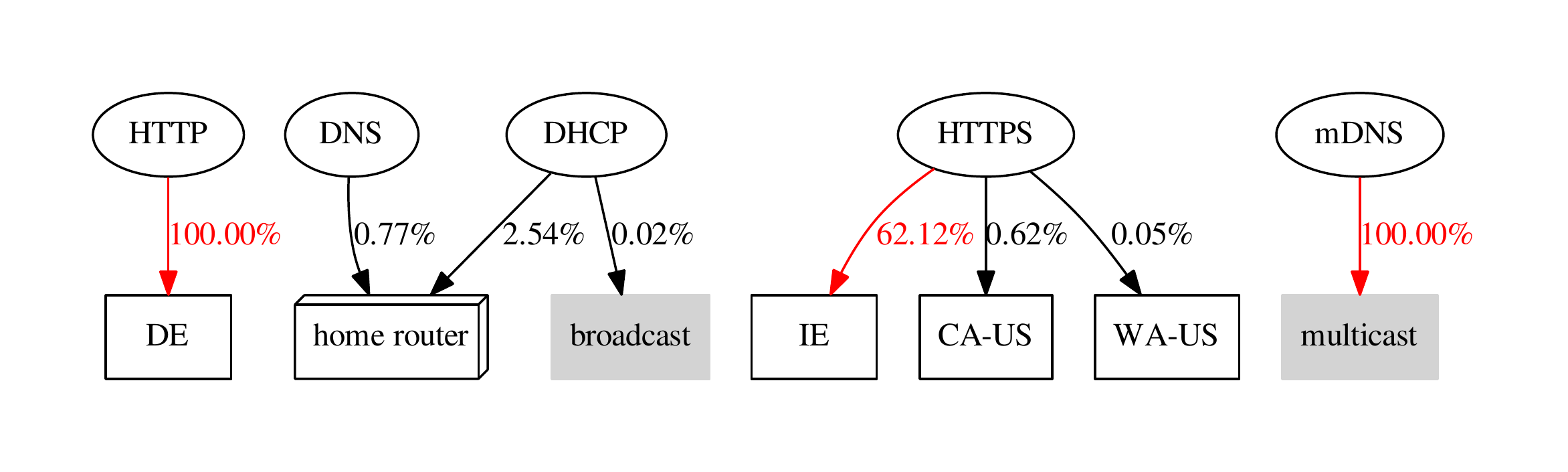}
  }
  \\
  \subfloat[Foobot]{
    \label{fig:foobot_geo}
    \includegraphics[height=22mm]{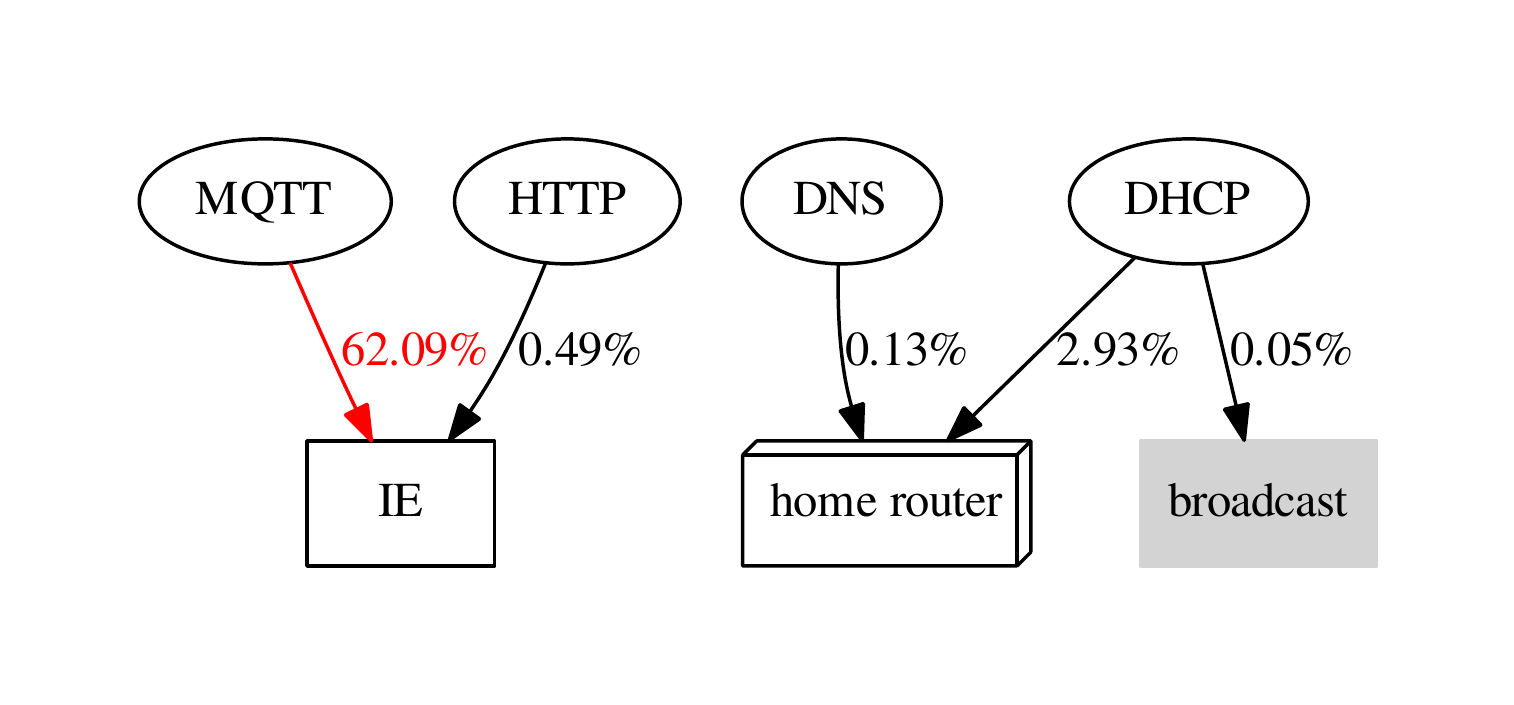}
  }
  \subfloat[Smart Plug]{
    \label{fig:smartplug_geo}
    \includegraphics[height=22mm]{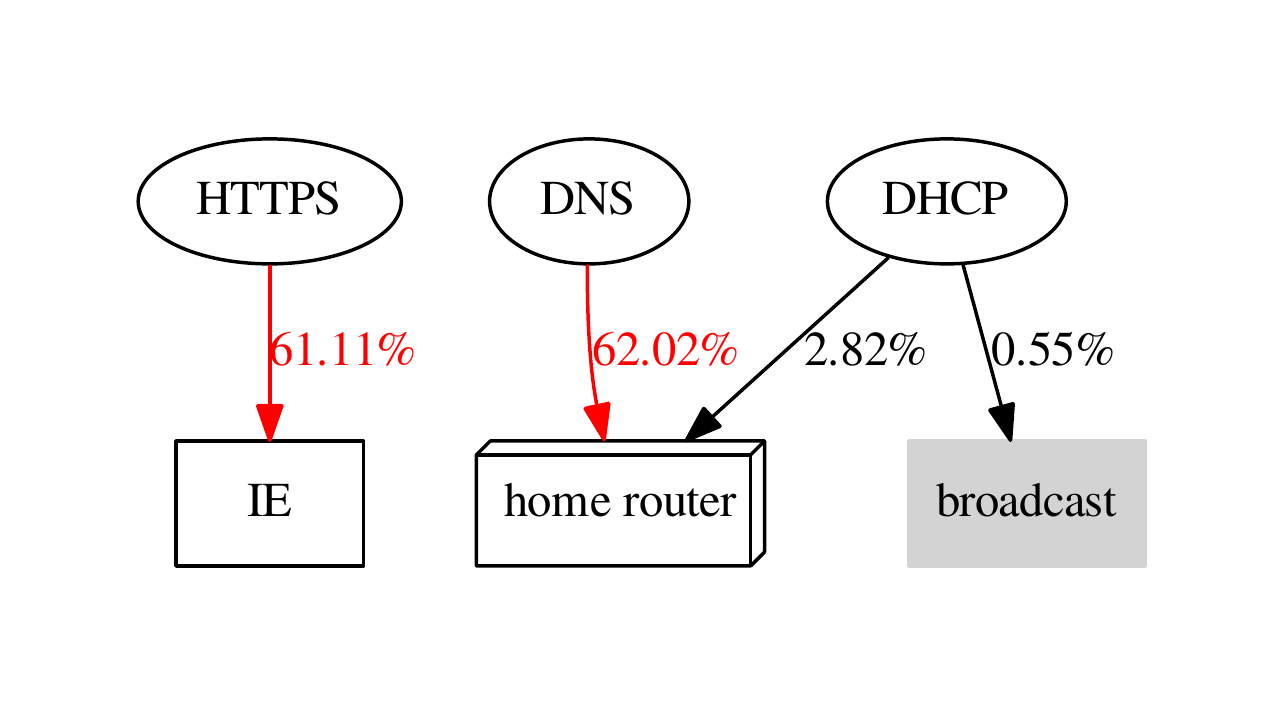}
  }
  \\
  \subfloat[Google Home]{
    \label{fig:googlehome_geo}
    \includegraphics[height=22mm]{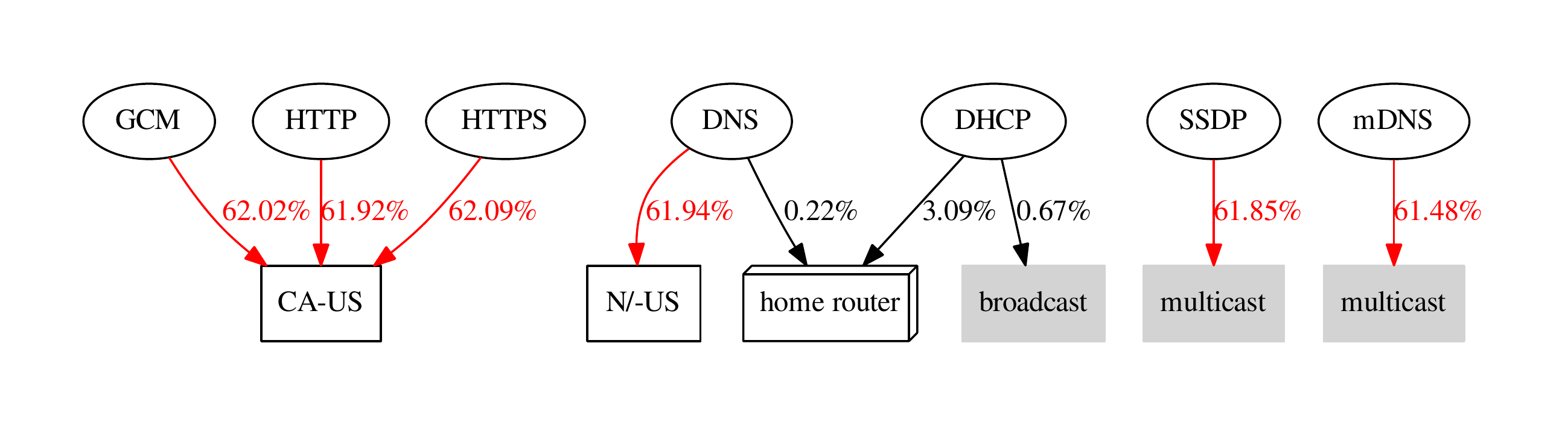}
  }
  \\
  \subfloat[Amazon Echo]{
    \label{fig:echo_geo}
    \includegraphics[width=\columnwidth]{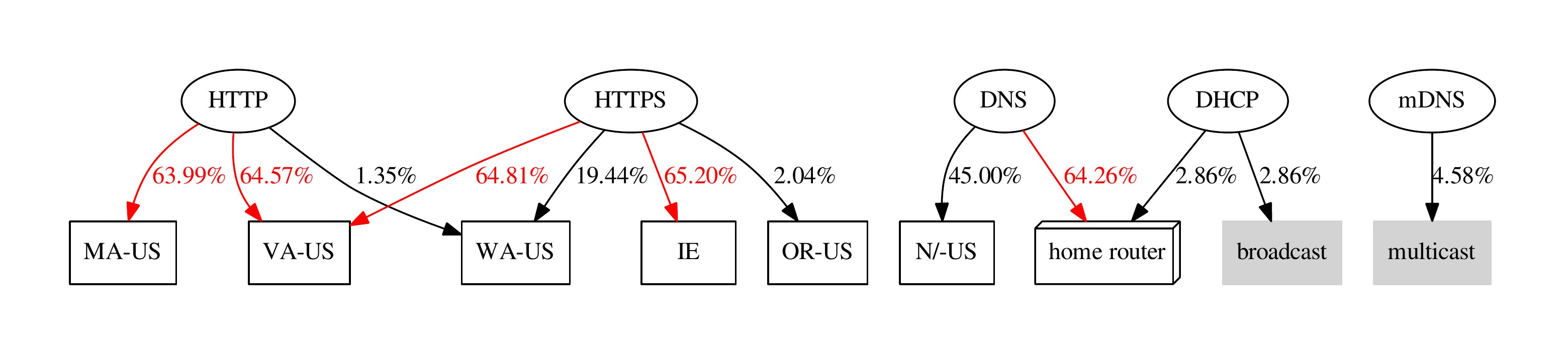}
  }
  \\
  \subfloat[Security Camera Hub]{
    \label{fig:camera_geo}
    \includegraphics[height=22mm]{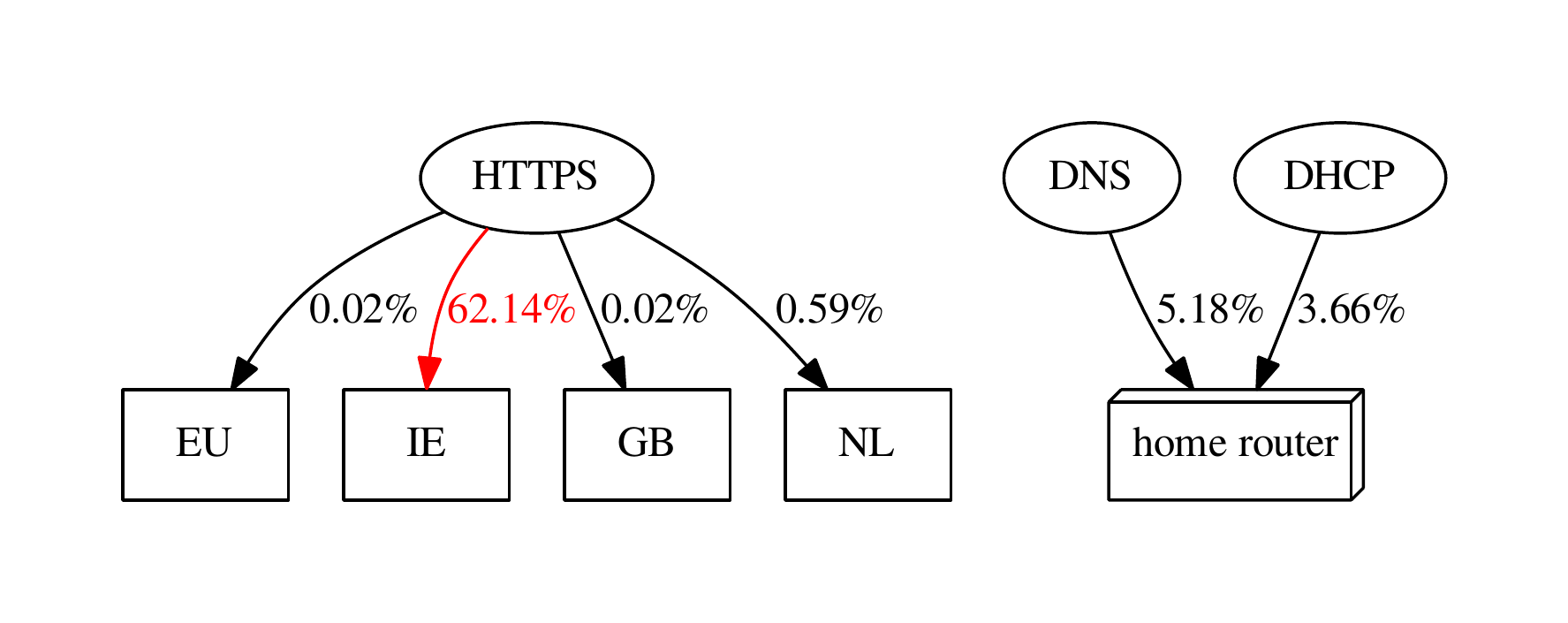}
  }
  \\
  \subfloat[Nest Smoke Alarm]{
    \label{fig:nestsmoke_geo}
    \includegraphics[height=22mm]{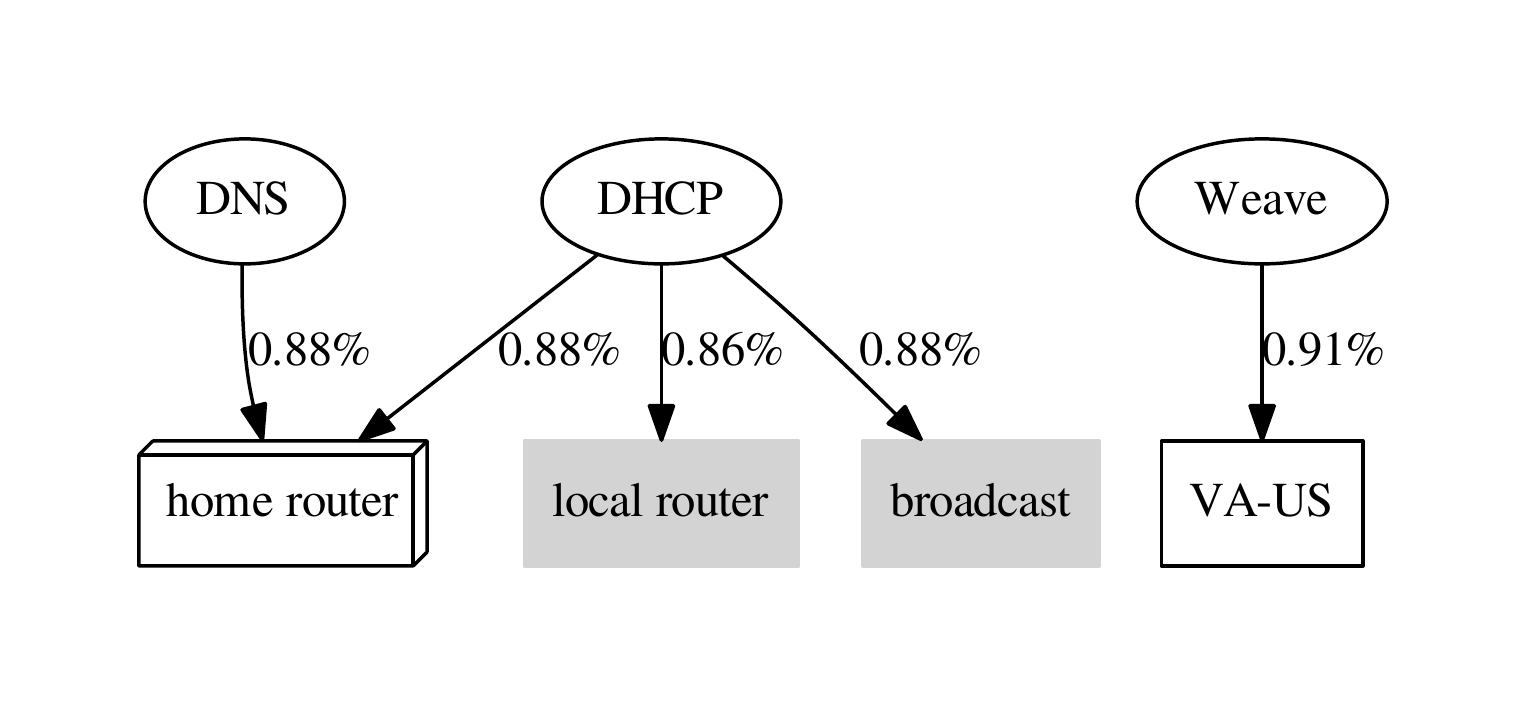}
  }
  \\
  \subfloat[D-Link Motion]{
    \label{fig:dlinkmotion_geo}
    \includegraphics[height=22mm]{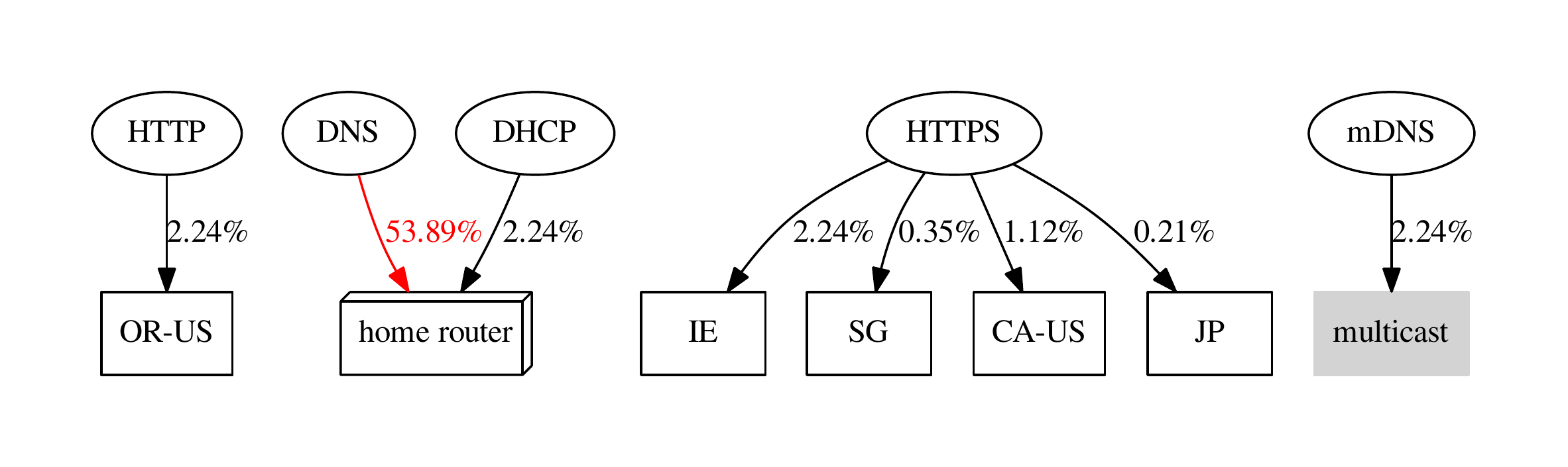}
  }
  \caption{\label{geo_series_1}Geolocated service accesses by device.}
\end{figure}

Finally, we performed IP geolocation (using IP address allocation and routing data to infer the geographical location of a host with a particular IP address) to estimate the different countries hosting the services used by our devices. We used a command line utility on Linux platform, \textit{geoiplookup}, to get the geolocation of a given IP address. For each combination of IoT device and application layer service, we aggregate the country names (also state names when the country is the USA) based on the ensemble of the IP addresses that the device communicates with. We filter out the NTP dependencies for Amazon Echo and Smart Plug as they each accessed tens of different locations due to their unusual usage pattern for NTP and DNS. We also extract some preliminary temporal information to describe the relative activeness of the communication: we divide the trace for each device into 15-minute windows and calculate the percentage of \emph{active windows}, those where communication to/from a specific location actually happens. The results are shown in Figure~\ref{geo_series_1}.


\begin{figure*}
  \centering
  \subfloat[Smart Plug]{
    \label{fig:smartplug_ntp_geo}
    \includegraphics[width=0.5\textwidth]{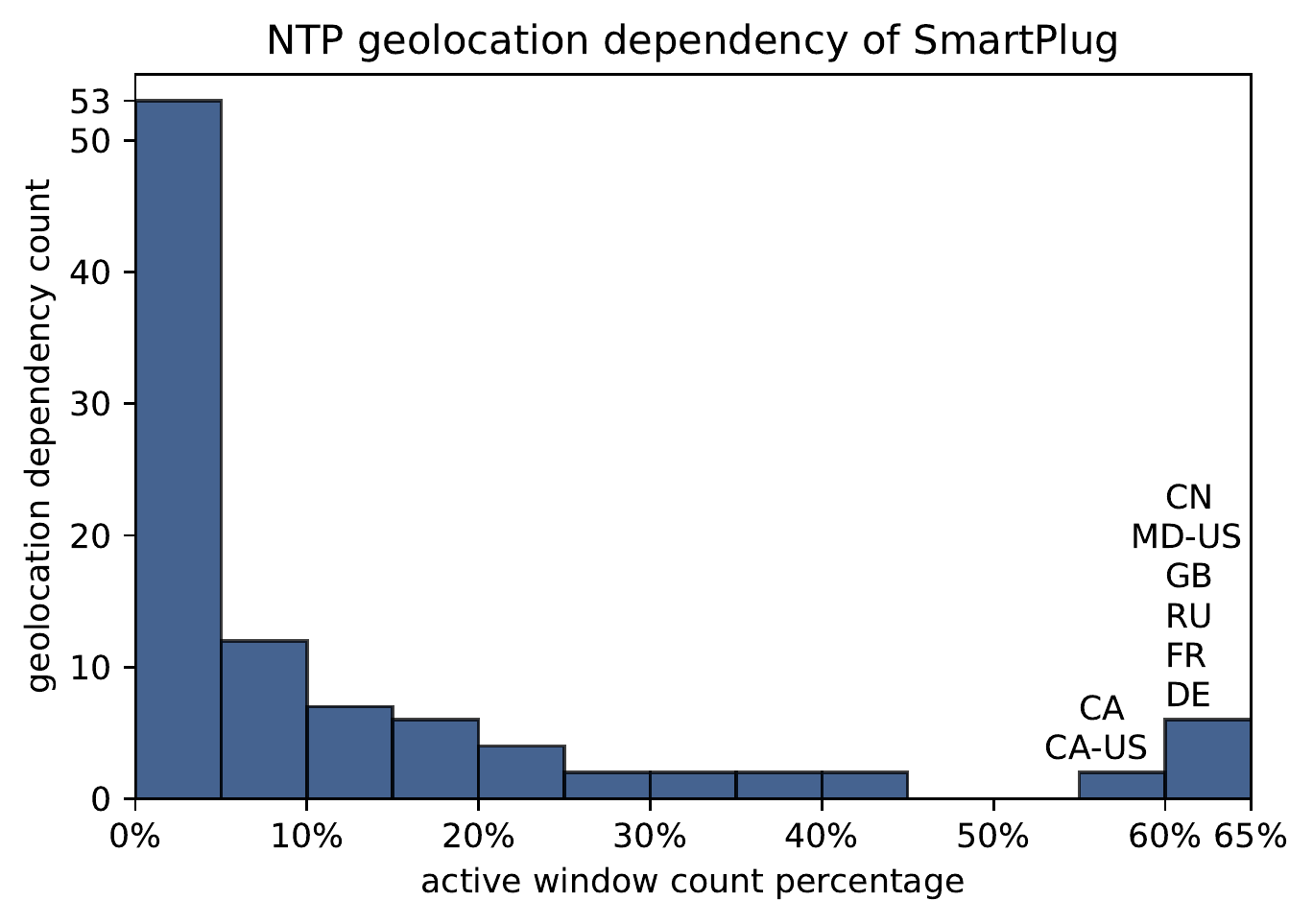}
  }
  \subfloat[Amazon Echo Dot]{
    \label{fig:echo_ntp_geo}
    \includegraphics[width=0.5\textwidth]{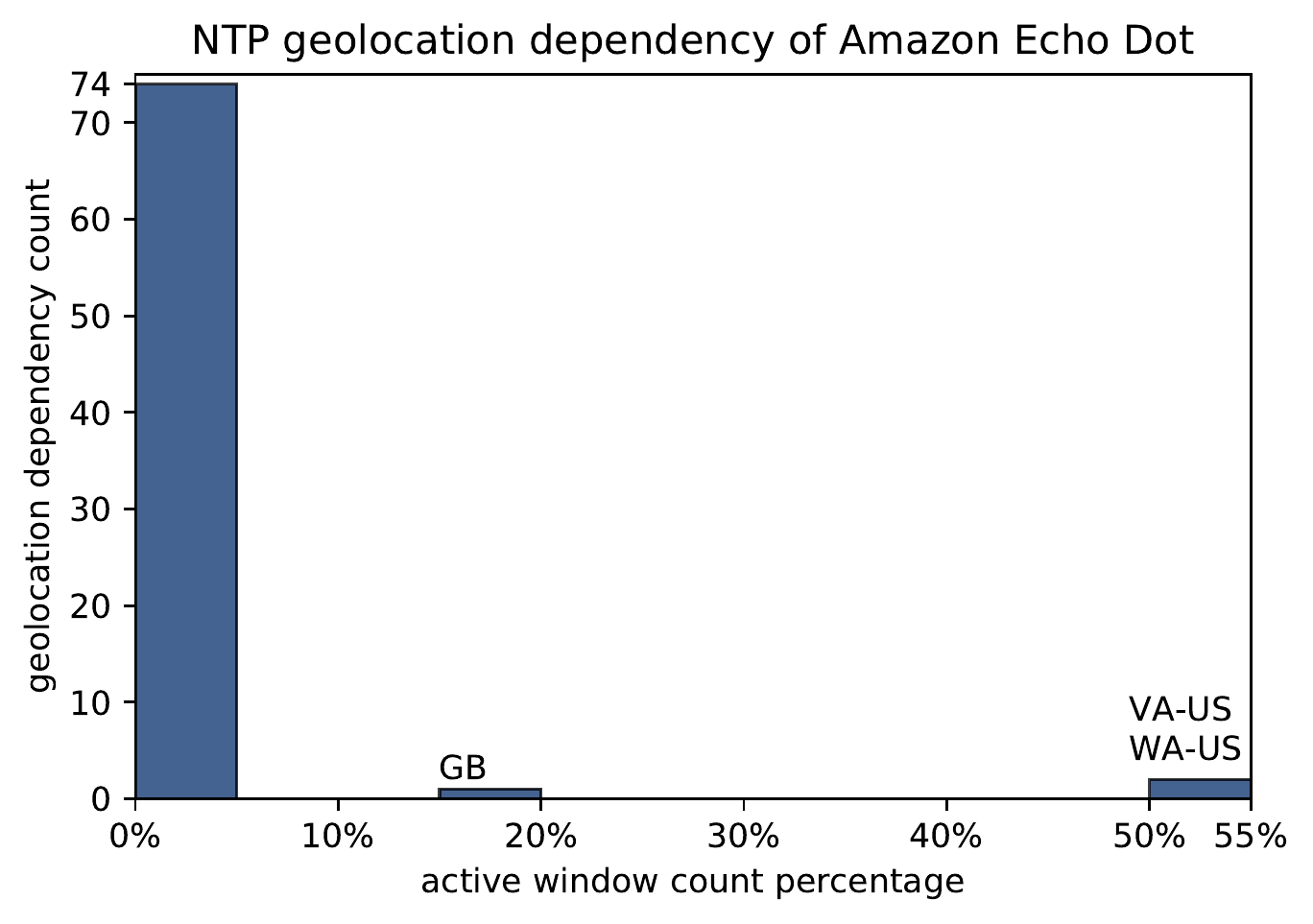}
  }
  \caption{\label{fig:geo_series_2}Summary of locations accessed by the NTP
    services.}
\end{figure*}

Both the TP-Link Smart Plug and D-Link Motion Sensor make a large number of DNS queries to global NTP servers (Figure~\ref{fig:geo_series_2}). The total DNS traffic generated by just 4 devices makes nearly 12\% of \emph{total} traffic generated from our setup. On the other hand, Hive Hub and Foobot air quality monitor make very few DNS queries to their servers. As we can see, most locations are quite inactive, with fewer than 10 locations where $>$50\% windows are active.

\section{Effects of Network Disruption}
\label{s:robust}

\begin{table}
  \begin{tabulary}{\textwidth}{R L c c c}
    \midrule
        & State                 & Internet & Local router & Devices\\
    \midrule
    \#1 & Steady state          & On       & On           & On \\
    \#2 & Internet disconnected & Off      & On           & On \\
    \#3 & Internet resumed      & On       & On           & On \\
    \#4 & Router power-off      & Off      & Off          & On \\
    \#5 & Router power-on       & Off      & On           & On \\
    \#6 & Internet resumed      & On       & On           & On \\
    \#7 & Device restarted      & On       & On           & Off$\rightarrow$On\\
    \midrule
  \end{tabulary}
  \caption{\label{tab:robustness}Experimental states use to examine effects of network disruption on devices.}
\end{table}

Table~\ref{tab:robustness} identifies seven different configurations for our experimental setup, each with different disruptions to device connectivity. We start with network and devices in steady state \st{1}, with all devices powered on, connected to the local router and thus to the Internet.

In our first experiment, we examine transitions from \st{1} $\rightarrow$ \st{2} $\rightarrow$ \st{3}, simulating Internet service interruption and recovery. We first cut off the Internet connection at the router by unplugging it from the University's network, and leave it unplugged for an hour to stabilise. In the meantime, we run scripts on the router to capture traces from the local network. We then reconnect the Internet connection to the router, and collect packet traces for a further hour.

\begin{figure*}
  \centering
  \subfloat[\st{1} $\rightarrow$ \st{2}]{
    \label{trans_summary_off}
    \includegraphics[width=.5\columnwidth]{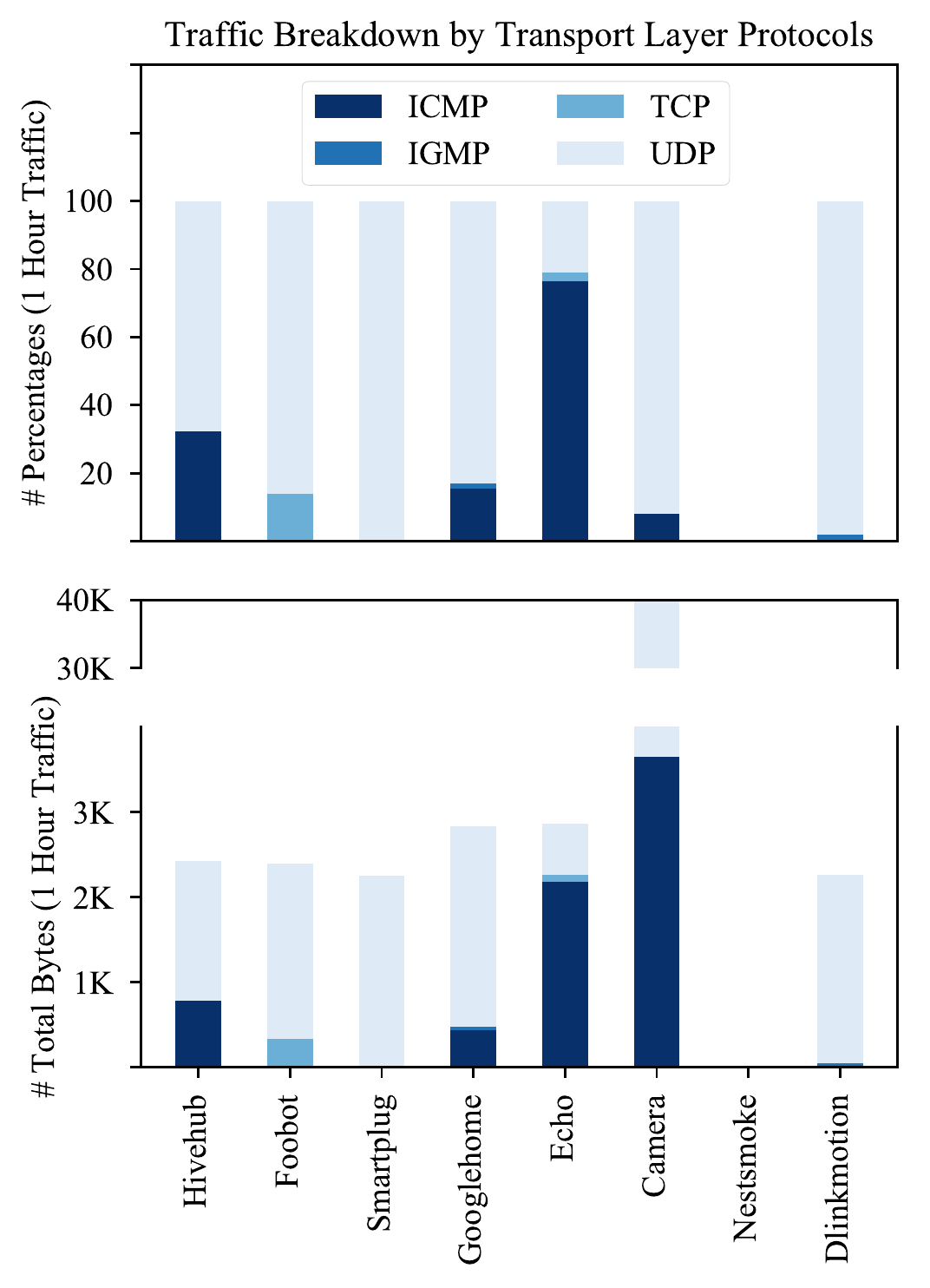}
  }
  \subfloat[\st{2} $\rightarrow$ \st{3}]{
    \label{trans_summary_on}
    \includegraphics[width=.5\columnwidth]{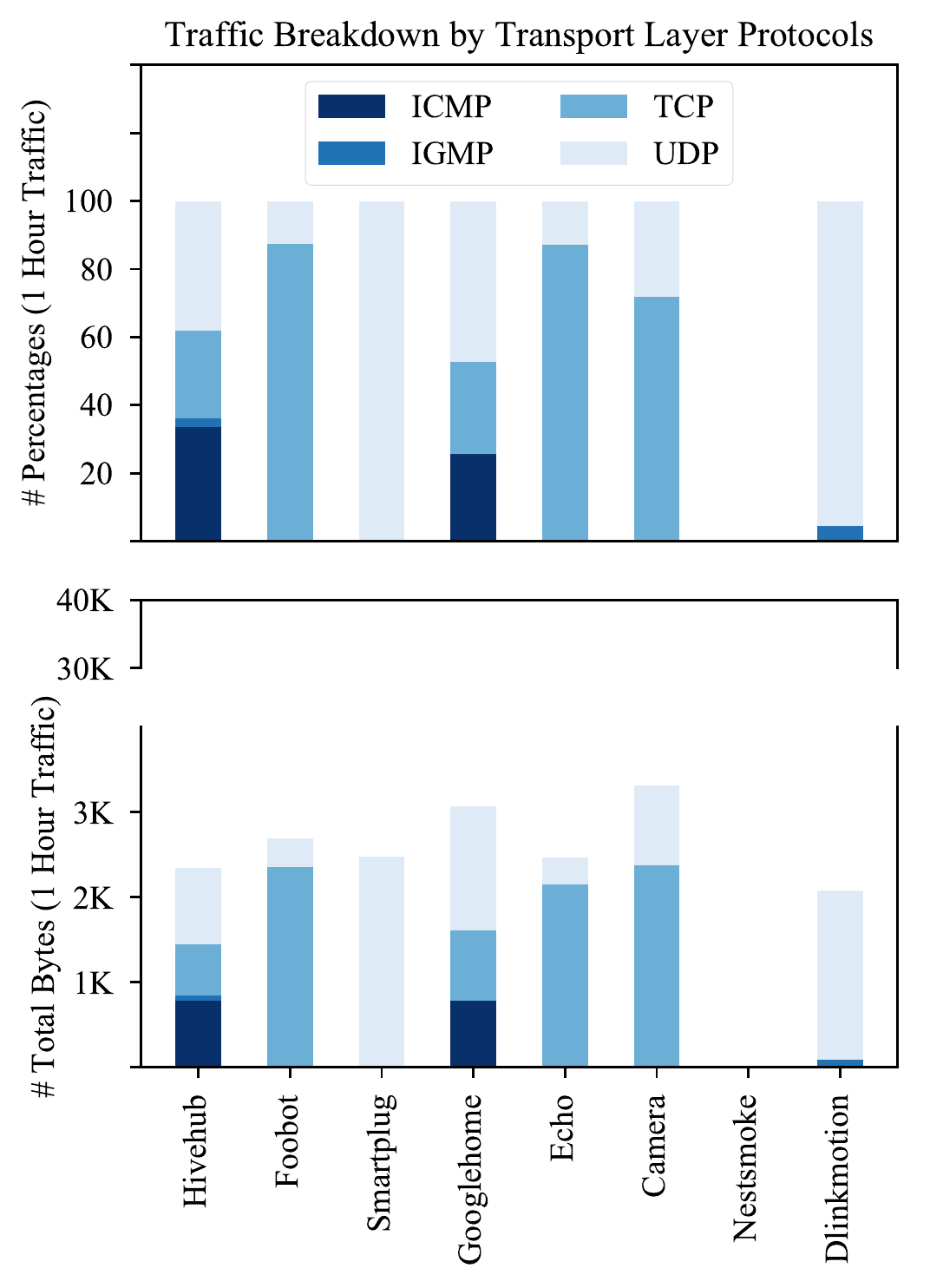}
  }
  \subfloat[\st{1} $\rightarrow$ \st{2}]{
    \label{serv_summary_off}
    \includegraphics[width=.5\columnwidth]{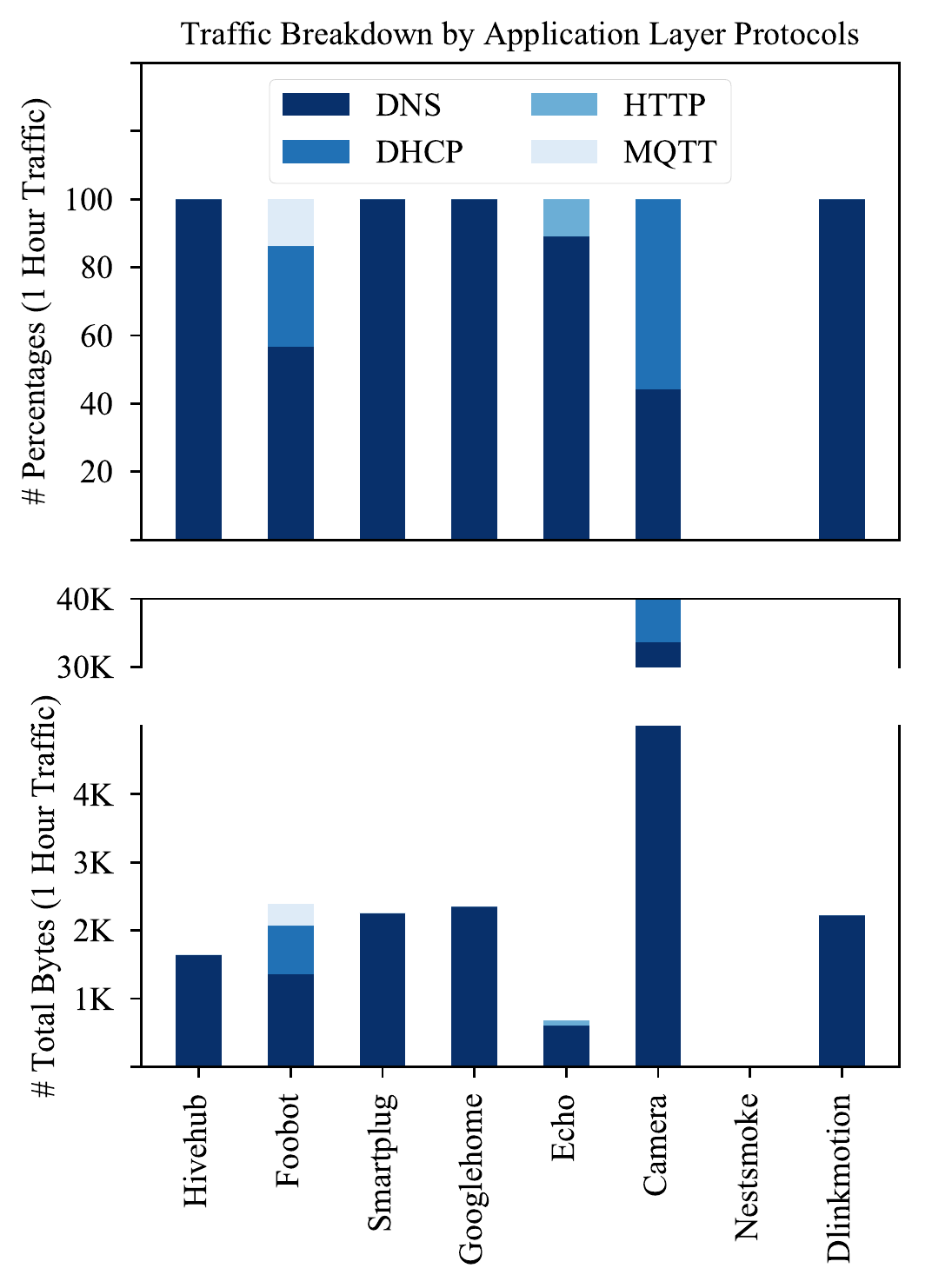}
  }
  \subfloat[\st{2} $\rightarrow$ \st{3}]{
    \label{serv_summary_on}
    \includegraphics[width=.5\columnwidth]{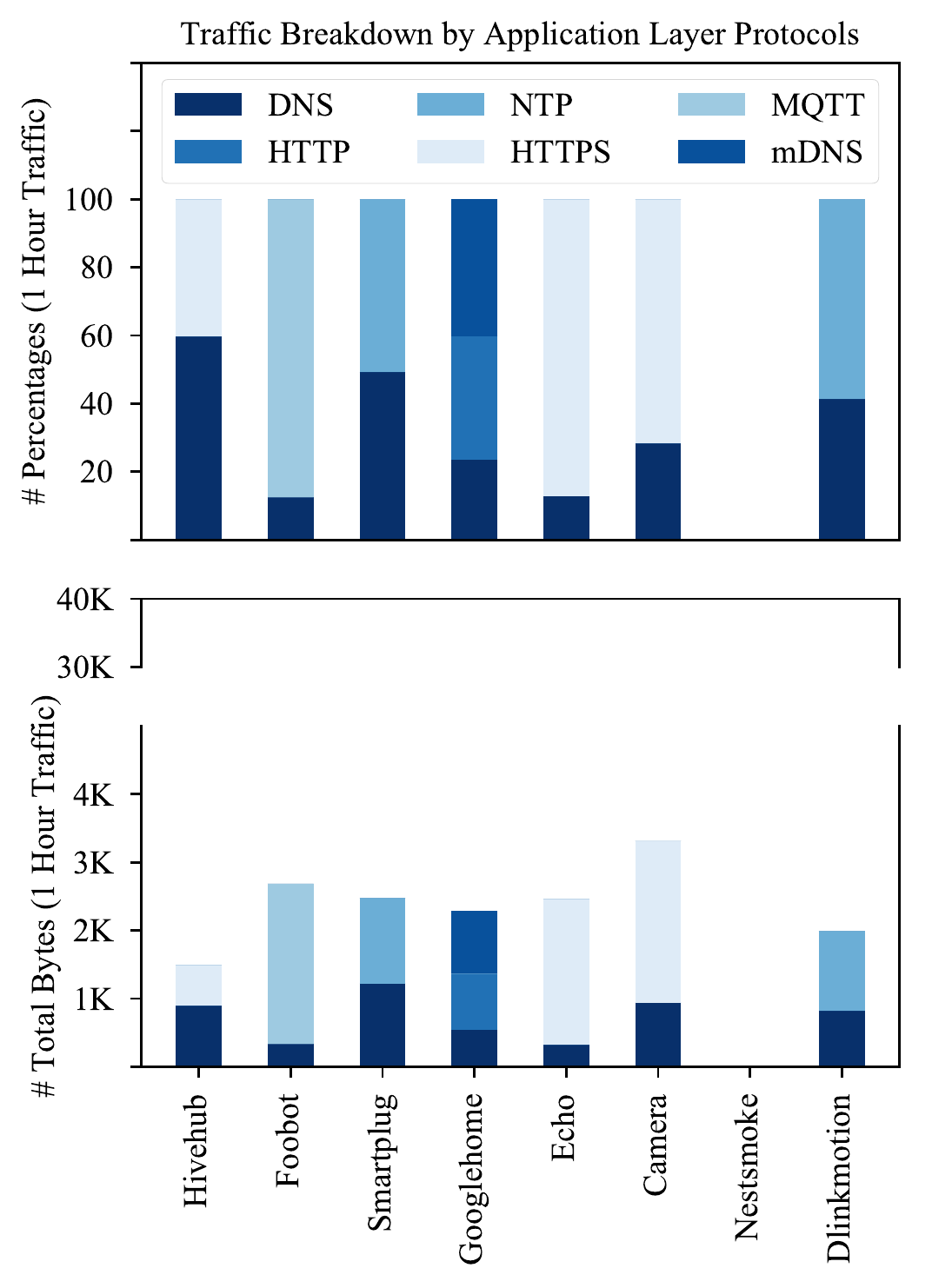}
  }

  \caption{\label{ts_offon} Traffic breakdown by transport- and application-layer Protocols for each device when network settings shown in Table~\ref{tab:robustness} are changed from the state \st{1} $\rightarrow$ \st{2}, and \st{2} $\rightarrow$ \st{3}. }
\end{figure*}

Figure~\ref{trans_summary} shows the transport-layer traffic generated in steady state \st{1}, while Figures~\ref{trans_summary_off} show the equivalent after disconnecting the Internet \st{2}. After disconnection there is a significant increase in UDP traffic for all devices, while TCP traffic is reduced to zero for all except the Foobot. The UDP traffic is mostly composed of DNS queries. The Security Camera Hub significantly increased the amount of DNS ($\sim$30 kB) and DHCP ($\sim$10 kB) traffic in the following hour. All hub devices (Hive Hub, Google Home, Echo, and Camera Hub) also significantly increased their ICMP traffic, presumably attempting to diagnose the interruption and perhaps recover quickly when service is resumed; sensor devices continue to emit negligible ICMP traffic. Foobot and Echo are the only devices which transmit TCP traffic to the local router, Foobot's composed of MQTT and Echo's of HTTP. The Nest Smoke Alarm neither sent nor received packets in this interval and so failed to detect lack of Internet connectivity.

Figures~\ref{serv_summary_off} shows that only four application-layer protocols of the ten active in steady state were operational while there was no Internet connectivity \st{2}. During this period, only the TP-Link Smart Plug and Nest Smoke Alarm functioned as usual. Amazon Echo Dot and Google Home Mini responded to their wake words, but could only direct the user to check Internet connectivities. For all other devices, their respective apps showed them to be off-line, requiring user intervention. No devices seemed able to send HTTPS or HTTP traffic, severely limiting their functionality.

We then observe what happens when Internet connectivity resumes and we transition from \st{2} to \st{3}. Figures~\ref{trans_summary_on} and~\ref{serv_summary_on} show the traffic generated when resuming. We see that only two of the four hubs (Hive Hub and Google Home) continue sending ICMP traffic, and only two devices (Hive Hub and D-Link Motion sensor) keep issuing IGMP requests. Five of the eight devices sent some TCP traffic in this state, suggesting that they detected connectivity had recovered. The Smart Plug and D-Link Motion sensors did not start sending any TCP but sent only DNS and NTP traffic, perhaps indicating both devices need to update global time before any TCP connection could be made. All devices continued to send UDP traffic, primarily DNS queries but the frequency reduced by $>50\%$ compared to \st{2}.

Google Home only sent traffic composed of three Application layer protocols (MQTT, HTTP and DNS), showing that Google Home remains partially functional in \st{3} for at least an hour. Echo and Camera Hub showed similar behaviour, both devices stopped ICMP traffic and sent DNS and HTTPS traffic. The DNS traffic of Camera Hub is reduced to nearly $\sim$1\,kB from 35\,kB per hour.

\begin{table*}
  \centering
  \begin{tabulary}{\textwidth}{
      L L L L L
    }
    \midrule
      & Device
      & Functionality
      & Services Continued
      & Services Disrupted \\

    \midrule
    1 & Hive Starter Kit Hub~\cite{Hive2018}
      & Partial
      & DNS
      & DHCP, HTTPS \\

    2 & Foobot Air Quality Monitor~\cite{Foobot2018}
      & Partial
      & DNS, MQTT, DHCP
      & HTTP \\

    3 & TP-Link Smart Plug~\cite{Tp2018}
      & Partial
      & DNS
      & NTP, HTTPS, DHCP \\

    4 & Google Home Mini~\cite{Ghome2018}
      & Partial
      & DNS
      & HTTP, NTP, HTTPS, SSDP, GCM, mDNS\\

    5 & Amazon Echo Dot~\cite{Echo2018}
      & Partial
      & DNS, HTTP
      & NTP, HTTPS, DHCP, mDNS \\

    6 & Arlo Security Camera Hub~\cite{Arlo2018}
      & Partial
      & DNS, DHCP
      & NTP, HTTPS \\

    7 & Nest Smoke Alarm~\cite{Nest2018}
      & Partial
      & DNS
      & Weave, DHCP\\

    8 & D-Link Motion Sensor~\cite{Dlink2018}
      & Partial
      & DNS
      & NTP, DHCP, HTTP, HTTPS, mDNS \\

    \midrule
  \end{tabulary}
  \caption{\label{tab:DeviceDependency}Observed dependencies.}
\end{table*}

Figure~\ref{ts_power_offon} shows the results of our second experiment where we examine network transition from states \st{3} $\rightarrow$ \st{4} $\rightarrow$ \st{5} $\rightarrow$ \st{6} per Table~\ref{tab:robustness}: having removed Internet connectivity, we then power off the router \st{4} and test functionality of the two surviving devices from the first experiment. As expected, the Smart Plug's app works whether or not Internet connectivity is available as it only relies on local connectivity, but Nest Smoke Alarm's mobile app shows nothing about this disrupted connectivity as the user can still trigger an alarm check on the device through its Bluetooth connection. After the check, we turn the router's power back on, \st{4} $\rightarrow$ \st{5}. We schedule a script to start collecting network traces right after the system finishes booting. We again remain in that state for an hour before turning the Internet back on at the router, \st{5} $\rightarrow$ \st{6}. We summarise Internet dependency of the devices in Table~\ref{tab:DeviceDependency}.

\begin{figure}
  \centering
  \subfloat[Transport-layer traffic]{
    \label{trans_summary_power_off}
    \includegraphics[trim={0 0 0 .75cm},clip, width=.5\columnwidth]
    {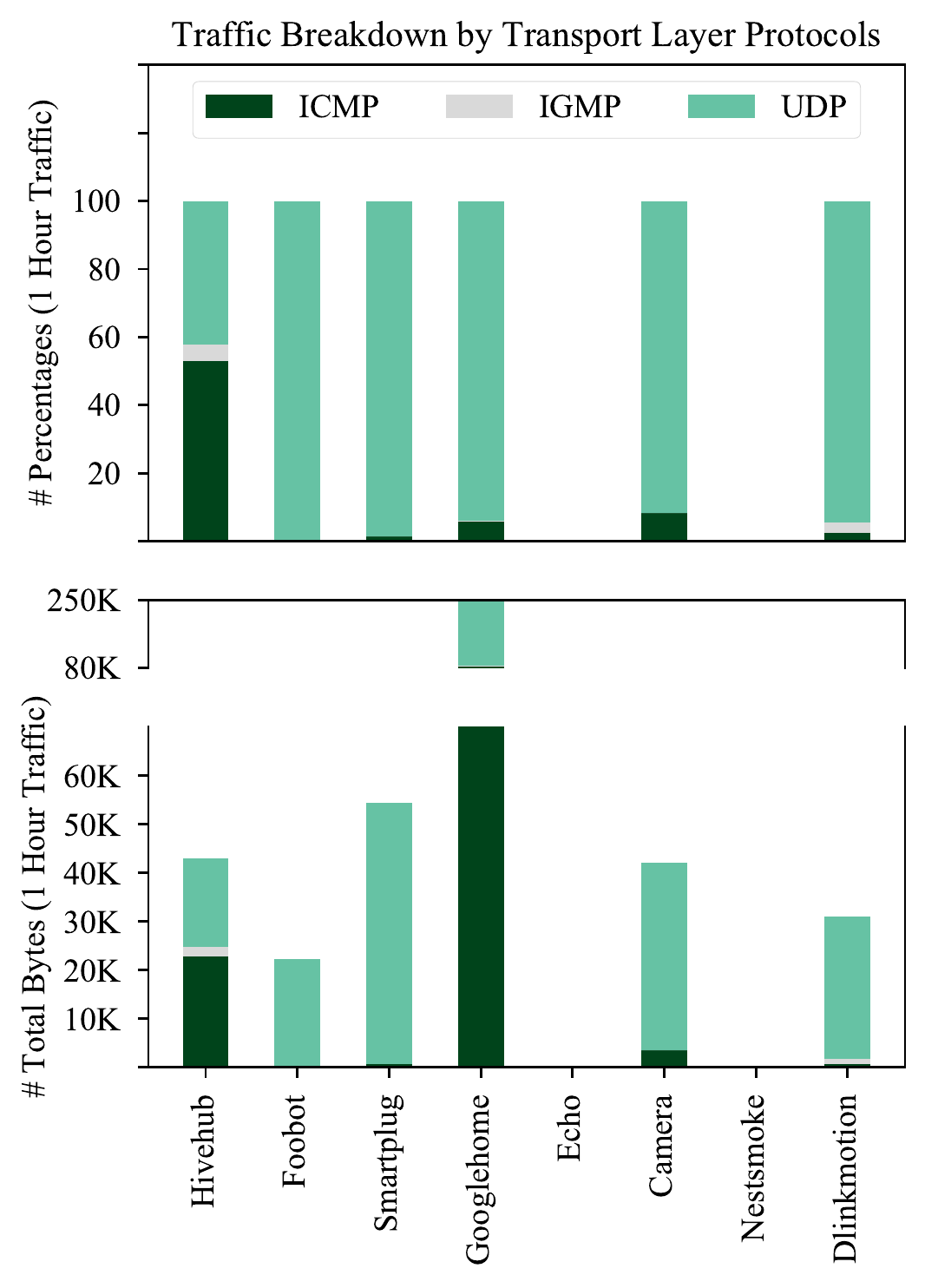}
  }
  ~
  \subfloat[Application-layer traffic]{
    \label{serv_summary_power_off}
    \includegraphics[trim={0 0 0 .75cm},clip, width=.5\columnwidth]
    {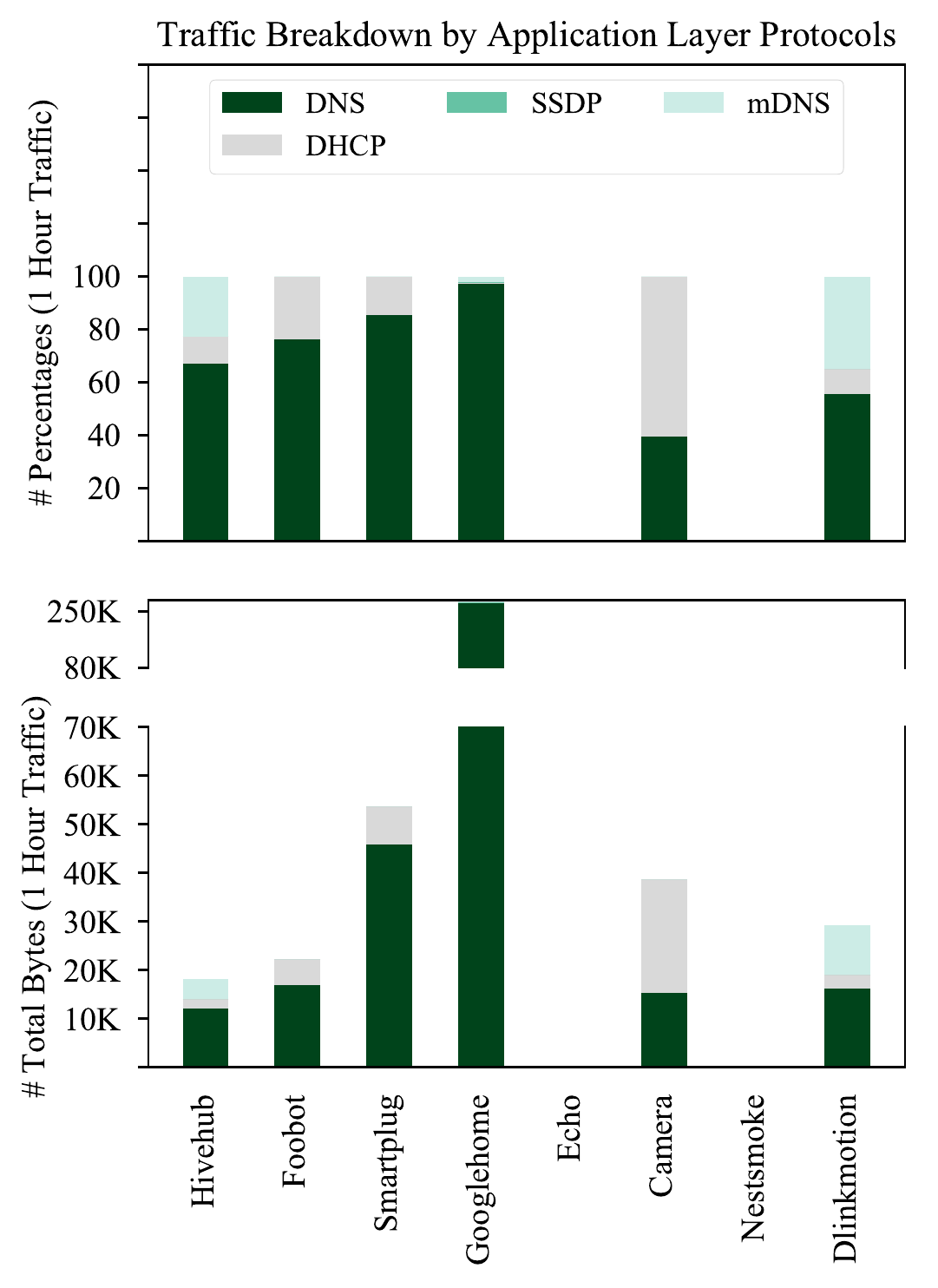}
  }
  \caption{\label{ts_power_offon} Traffic breakdown by transport- and application-layer protocols for each device when network settings shown in Table~\ref{tab:robustness} are transition states, \st{4} $\rightarrow$ \st{5}. }
\end{figure}

Figures~\ref{trans_summary_power_off} and \ref{serv_summary_power_off} show the effect when the router is powered on but Internet is still disconnected, i.e.,~\st{5}. We see that only three transport-layer protocols (ICMP, IGMP and UDP) generate significant traffic in that hour, and there was no TCP traffic. All devices generated significant UDP traffic ($>20$\,kB in first hour) as compared to \st{2} which suggests current state behaviour depends on previous state. Google Home generated $>$250\,kB of ICMP and UDP traffic with UDP traffic composed of DNS and SSDP. We found Echo behaved strangely after the router restart as it didn't detect the router automatically, instead connecting to our institution's open Wi-Fi network, resulting in missed traces from Echo.

\begin{figure}
  \centering
  \subfloat[Transport-layer traffic]{
    \label{trans_summary_power_on}
    \includegraphics[trim={0 0 0 .75cm},clip, width=.5\columnwidth]
    {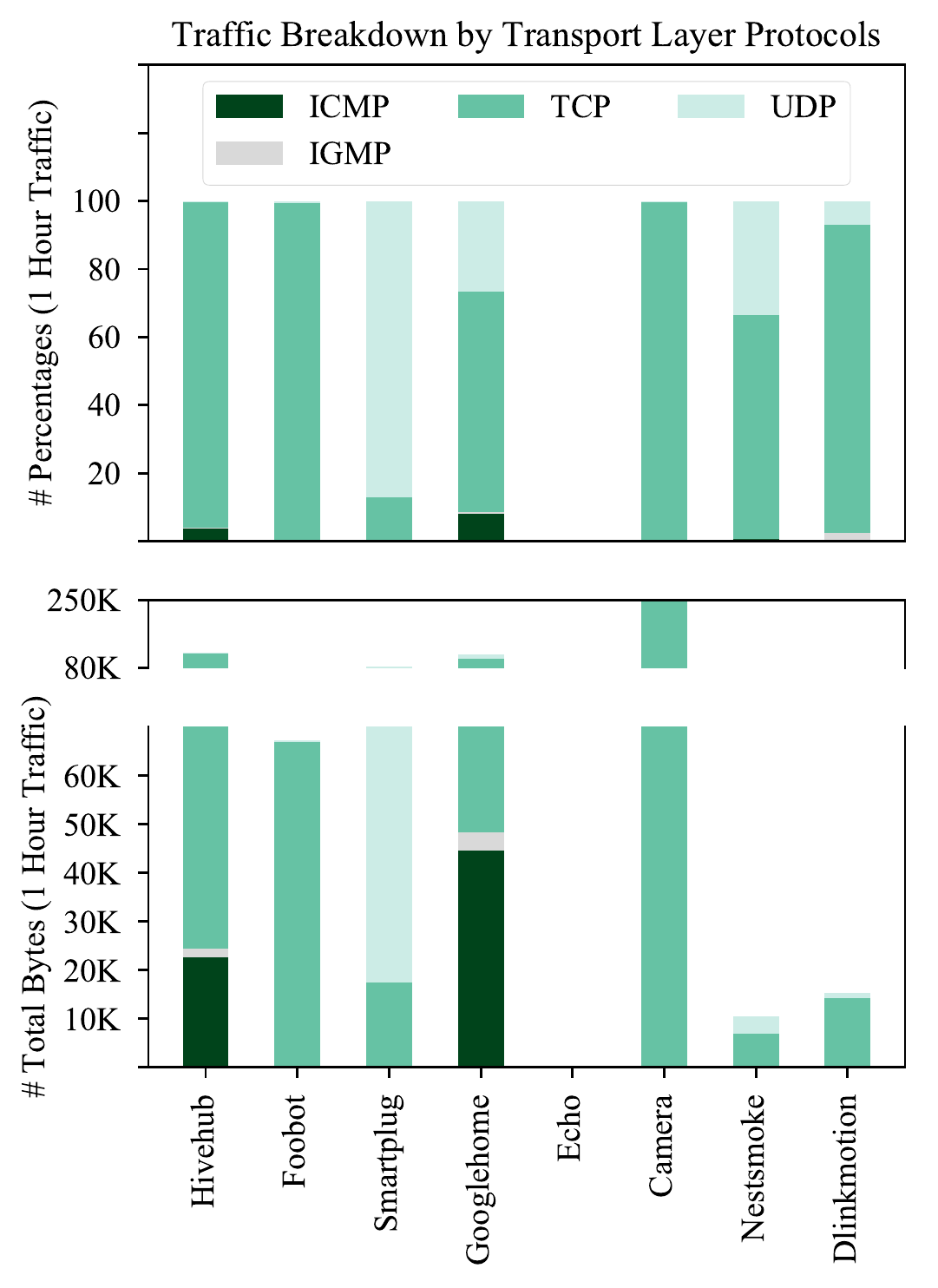}
  }
  ~
  \subfloat[Application-layer traffic]{
    \label{serv_summary_power_on}
    \includegraphics[trim={0 0 0 .75cm},clip, width=.5\columnwidth]
    {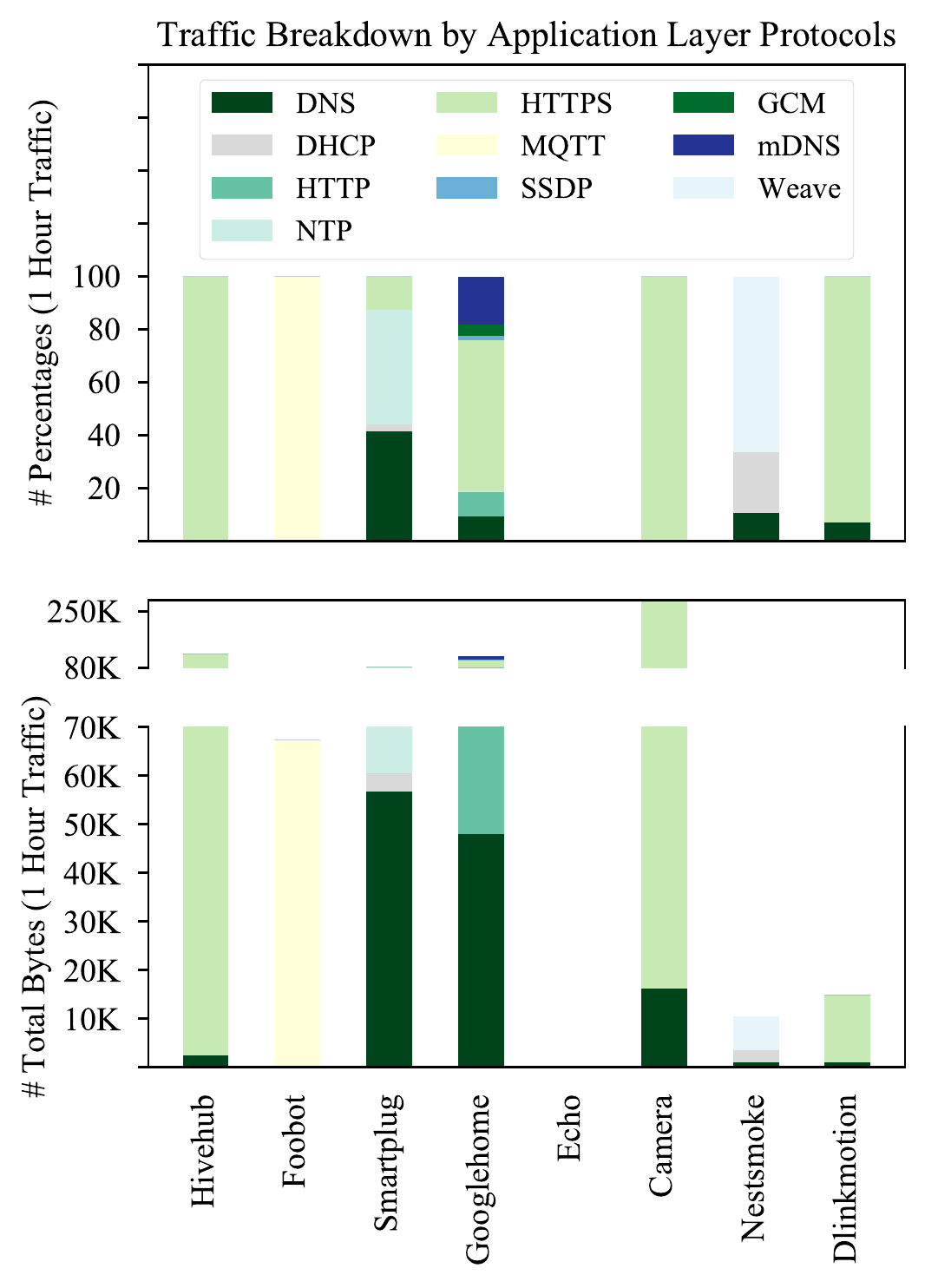}
  }

  \caption{\label{ts_power_offon} Traffic breakdown by transport- and application-layer protocols for each device when network settings shown in Table~\ref{tab:robustness} are transition states, \st{5} $\rightarrow$ \st{6}. }
\end{figure}

The effect of the transition from \st{5} $\rightarrow$ \st{6} is shown in
Figures~\ref{trans_summary_power_on} and \ref{serv_summary_power_on} where
Internet connectivity is restored at the local router. We find that all ten
application-layer protocols exhibit traffic, and all devices resume their normal
functionality. Smart Plug ($\sim$55\,kB), Google Home ($\sim$45\,kB) and Camera
Hub ($\sim$15\,kB) sent significant DNS traffic in the first hour. The magnitude
of total traffic generated by devices is nearly 10 times more than in state
\st{3}, even though both states are superficially similar. This suggests that local router restart does briefly create a measurable increase in IoT traffic.

\begin{figure}
  \centering
  \subfloat[Transport-layer traffic]{
    \label{trans_summary_restart_5m}
    \includegraphics[trim={0 0 0 .75cm},clip, width=.5\columnwidth]
    {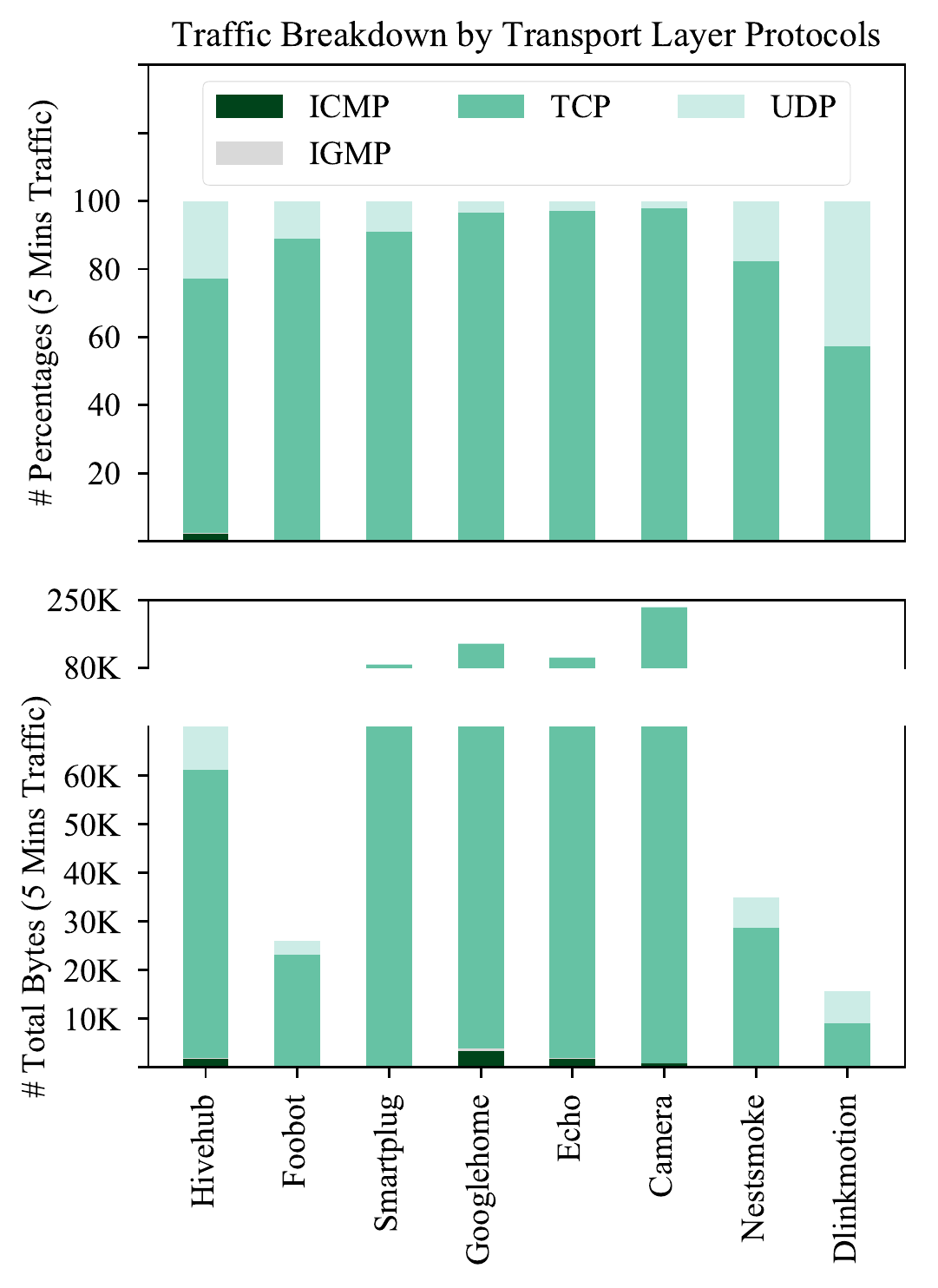}
  }
  ~
  \subfloat[Application-layer traffic]{
    \label{serv_summary_restart_5m}
    \includegraphics[trim={0 0 0 .75cm},clip, width=.5\columnwidth]
    {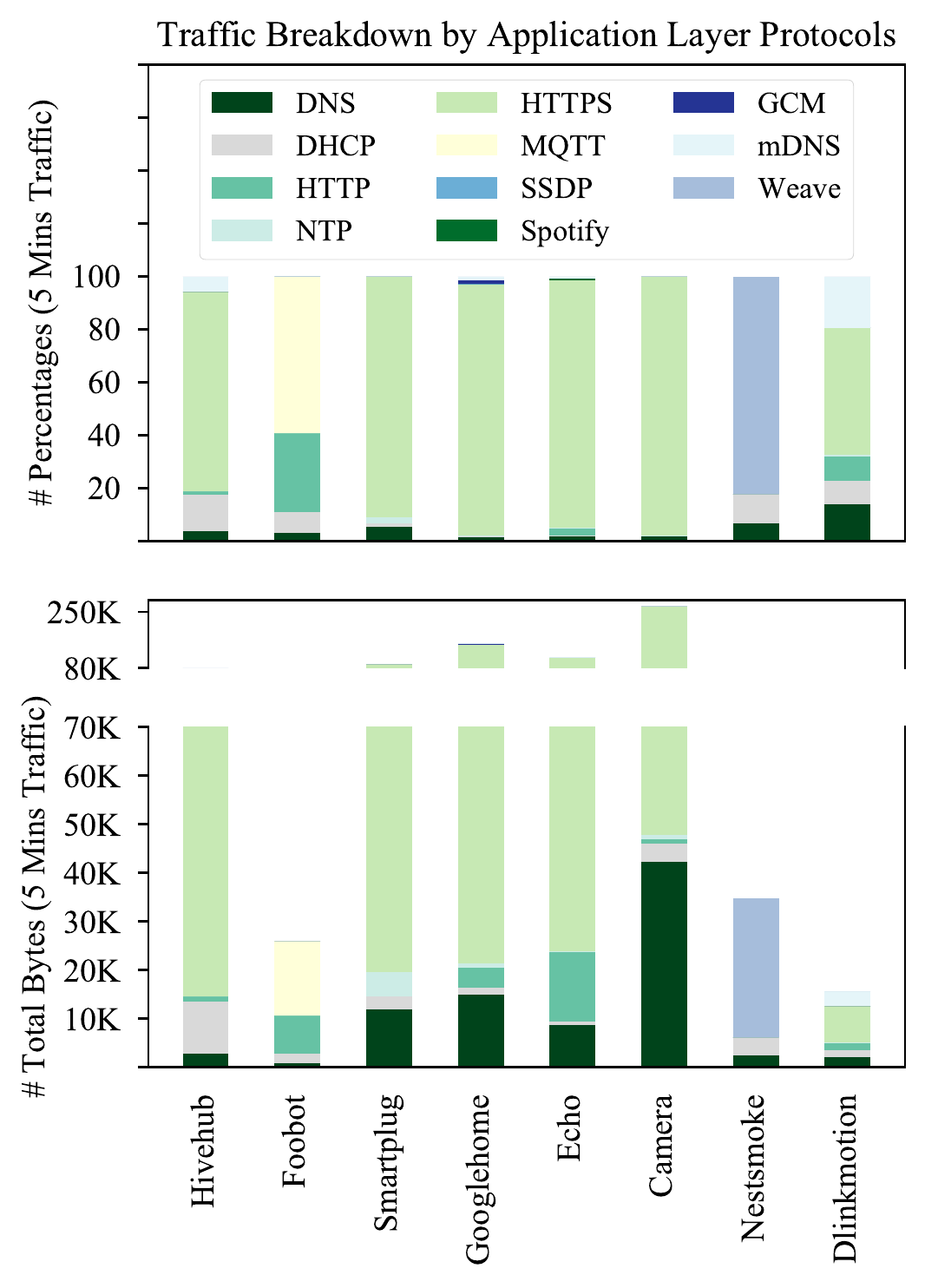}
  }
  \caption{\label{ts_restart_5m} Traffic breakdown by transport- and application-layer protocols for each device when network settings shown in Table~\ref{tab:robustness} are changed from the state \st{6} $\rightarrow$ \st{7}.}
\end{figure}

Figure~\ref{ts_restart_5m} summarises traffic in the first 5\,minutes after transitioning \st{6} $\rightarrow$ \st{7}. In this state, devices are restarted one by one while the router remains powered on and Internet connected. After each device is restarted, we carried out one activity with it to ensure it was functioning normally. Each device was then kept idle for the rest of the measurement time. We analysed traffic from the  5\,mins immediately after the device was restarted, and summarise this in Figure~\ref{ts_restart_5m}. To our surprise, all devices generated traffic equivalent to the traffic generated in state \st{6} (one hour) in that 5\,mins period. Compared to other previous states we observed the Echo traces to show port access queries -- both HTTP and HTTPS -- which were not previously observed. Also the Nest Smoke Alarm generated a total of $\sim$35\,kB of traffic in that 5\,min period after restart, suggesting device restart increases network traffic.


\section{Related Work}
\label{s:related}

The challenges posed by the current state of the IoT ecosystem are widespread, providing individual, technological and societal threats. In this context we define a {\em threat} as the danger resulting from exploitation of vulnerabilities in a system causing potential harm~\cite{Arias2015, Bugeja2017}. For example, Bugeja et. al~\cite{Bugeja2017} analyse the potential threat agents in home IOT environments and classify them based on their motivations and capabilities. The various threat agents include nation states, terrorists, competitors, and criminals. Their various motivations involve curiosity, personal gain, terrorism, and national interest. To minimise possible threats, deployments will need to provide for various security requirements: authentication, confidentiality, integrity, non-repudiation and availability~\cite{Aldosari2015, Ren2017}. Additionally, privacy risks and Human-Data Interaction challenges must be managed, with support for data subjects' rights to control, edit, manage and delete information about themselves, as well as deciding when, how and the extent to which information about them may be communicated to others~\cite{Westin1967,hdi}.

Recent years have seen both privacy and security perspectives explored, analysed and presented in many research articles, e.g.,~\cite{Arias2015, Bugeja2017}. However, there has been relatively little research on how the critical end-to-end services and infrastructure components of the IoT ecosystem could affect scalability, availability, and integrity of these systems. In response, we focus here on understanding the threats linked to the scalability, availability, and integrity that deployment of commodity IoT devices will create.

Abomhara et al~\cite{Abomhara2015} analyse IoT threat types and characterise intruders and attacks facing IoT devices and services. Connected devices are found to be rather valuable to cyber-attackers for several reasons: \one~most IoT devices operate unattended by humans, so it is easy for an attacker to gain physical access to them; \two~most IoT components communicate over wireless networks without requiring encryption, so attackers might obtain confidential information simply by eavesdropping; \three~most IoT components cannot support complex security schemes due to low power and computing resource capabilities. \footnote{Indeed, we observed this ourselves when we struggled to get one of our devices connected because our institution's wireless network policy mandated an enterprise-level security protocol while the device only supported the commonly deployed domestic security protocols.}

Others have recently carried out similar analysis and studies looking into various IoT traffic features, though most focus on privacy and security. Abomhara et al~\cite{Abomhara2015} analysed IoT threat types and characterise intruders and attacks facing IoT devices and services. Apthorpe et al~\cite{Apthorpe2018} discusses privacy leakage from DNS queries and encrypted traffic. Duo et al~\cite{Duo2017} consider what happens in different scenarios were volumetric data generated between user and cloud services from the compromised devices and app services is exposed. They present a machine learning mechanism to learn the pattern identifying a DDoS attack, install a corresponding filter at the edge of the network, and discuss the amplification factor when services are disrupted due to DDoS attacks.

Dainotti et al~\cite{Dainotti2016} discusses IoT network's packet level traffic analysis, inter-packet time and packet size. The analysis done in the paper provides details such as delay, jitter and packet loss. Mahjabin et al~\cite{Mahjabin2017} discusses impact and scale of DDoS attack. Apthorpe et al~\cite{Apthorpe2017a} suggests four strategies for protecting smart home privacy from home network observers, for example, blocking traffic, concealing DNS, tunnelling traffic, and reshaping traffic. Andradez et al~\cite{Andrade2017} studied the connection time of cars, both spatial and temporal data to find correlation using the time series representation.

Lee et al~\cite{Lee2014} discuss sequential correlation between DNS queries to show the temporal dependency and vulnerability. Allman et al~\cite{Allman2007} discuss issues and etiquette concerning the use of shared measurement data. Kumar et al~\cite{Kumar2018} discuss the mis-issuance of security certificate using Zlint, and found only 0.02\% of certificate are mis-issued. Lai et al~\cite{Lai2015} present an algorithm to visualise top DNS server queries. Krohnke et al~\cite{Krohnke2018} discuss the impact of location of DNS server on DNS query responses if, for example, it is within only one AS or domain. Hahn et al~\cite{Hahn2018} present interesting work on detection and separation of compressed text from encrypted text using $k$-nearest neighbour (60\%) and 1D convoluted neural network (66.9\%).

Finally, Sivanathan et al~\cite{sivanathan17:charac.class.iot.traff.smart.cities.campus} study IoT traces from a selection of commodity IoT devices. They observe some similar properties to those we report here, but focus analysis on active/inactive periods (finding that most active periods are short), and on clustering observed behaviours among their devices (finding approximately 5 clusters of network behaviour). They use these data and analyses to classify and ultimately identify devices. In contrast, we analyse behaviour in more detail (traffic volumes and periodicity by protocol), and we are particularly interested in the implications for resilience of the built environment with respect to the dependencies we implicitly take by widespread deployment of IoT devices through the network connectivity and Internet services they rely upon.

\section{Conclusions}
\label{s:conclusions}

The traffic capture and analysis presented here adds to the body of work presenting the network-level behaviour of commodity IoT devices. We add a protocol-level breakdown and more detailed analysis of periodicity over a longer time period. This leads to an exploration of the service and infrastructure dependencies that will be taken in ``smart'' environments when IoT devices are deployed. We thus also present analysis of some of these dependencies from a cloud service and geographical perspective, finding that many devices make use of services distributed across the planet and thus appear dependent on the global network infrastructure even when carrying out purely local actions. Finally, we examine the robustness of device operation when connectivity is disrupted, finding that some devices cease to operate properly without network connectivity (even where their behaviour appears, on the face of it, to require only local information, e.g.,~the Hive thermostat). Further, they exhibit quite different network behaviours, typically involving significantly more traffic and possibly use of otherwise unobserved protocols, when connectivity is recovered after some disruption. This has implications for device behaviour profiling and firewalling as proposed by, e.g.,~the IETF's draft Manufacturer Usage Description (MUD) standard~\cite{ietf-opsawg-mud-25}.

\begin{acks}
 This project was supported by a mini-projects award from the Centre for
 Digital Built Britain and Innovate UK under Grant 90066, and by EPSRC
 EP/N028260/1 and EP/R03351X/1.
\end{acks}

{
  \bibliographystyle{ACM-Reference-Format}
  \bibliography{iotdi19}
}
\end{document}